\documentclass[sigconf]{acmart}
\usepackage{tabularx}
\usepackage{algorithm}
\usepackage{algorithmic}
\usepackage{etoolbox}
\usepackage{xcolor}
\usepackage{threeparttable} 
\newtoggle{showadditions}
\newtoggle{showdeletions}
\toggletrue{showadditions}
\togglefalse{showdeletions}
\newcolumntype{C}[1]{>{\centering\let\newline\\\arraybackslash\hspace{0pt}}m{#1}}

\AtBeginDocument{%
  }

\copyrightyear{2025}
\acmYear{2025}
\setcopyright{cc}
\setcctype{by}
\acmConference[ASSETS '25]{The 27th International ACM SIGACCESS Conference
on Computers and Accessibility}{October 26--29, 2025}{Denver, CO, USA}
\acmBooktitle{The 27th International ACM SIGACCESS Conference on Computers
and Accessibility (ASSETS '25), October 26--29, 2025, Denver, CO, USA}
\acmDOI{10.1145/3663547.3746386}
\acmISBN{979-8-4007-0676-9/2025/10}

\begin{document}

\title[Video Customization for Viewers with ADHD]{FocusView: Understanding and Customizing Informational Video Watching Experiences for Viewers with ADHD}


\author{Hanxiu `Hazel' Zhu}
\affiliation{%
  \institution{University of Wisconsin-Madison}
  \city{Madison}
  \state{Wisconsin}
  \country{USA}
}
\email{hzhu339@wisc.edu}

\author{Ruijia Chen}
\affiliation{%
  \institution{University of Wisconsin-Madison}
  \city{Madison}
  \state{Wisconsin}
  \country{USA}
}
\email{ruijia.chen@wisc.edu}

\author{Yuhang Zhao}
\affiliation{%
  \institution{University of Wisconsin-Madison}
  \city{Madison}
  \state{Wisconsin}
  \country{USA}
  }
\email{yuhang.zhao@cs.wisc.edu}

\renewcommand{\shortauthors}{Zhu et al.}

\begin{abstract}
 While videos have become increasingly prevalent in delivering information across different educational and professional contexts, individuals with ADHD often face attention challenges when watching informational videos due to the dynamic, multimodal, yet potentially distracting video elements. To understand and address this critical challenge, we designed \textit{FocusView}, a video customization interface that allows viewers with ADHD to customize informational videos from different aspects. We evaluated FocusView with 12 participants with ADHD and found that FocusView significantly improved the viewability of videos by reducing distractions. Through the study, we uncovered participants' diverse perceptions of video distractions (e.g., background music as a distraction vs. stimulation boost) and their customization preferences, highlighting unique ADHD-relevant needs in designing video customization interfaces (e.g., reducing the number of options to avoid distraction caused by customization itself). We further derived design considerations for future video customization systems for the ADHD community.   

\end{abstract}

\ccsdesc[500]{Human-centered computing~Accessibility systems and tools}
\ccsdesc[500]{Human-centered computing~Accessibility design and evaluation methods}

\keywords{Individuals with Disabilities \& Assistive Technologies, Video Accessibility, ADHD}


% %%
% %% Keywords. The author(s) should pick words that accurately describe
% %% the work being presented. Separate the keywords with commas.
% \keywords{Do, Not, Us, This, Code, Put, the, Correct, Terms, for,
% %% A "teaser" image appears between the author and affiliation
% %% information and the body of the document, and typically spans the
\begin{teaserfigure}
    \includegraphics[width=\textwidth]{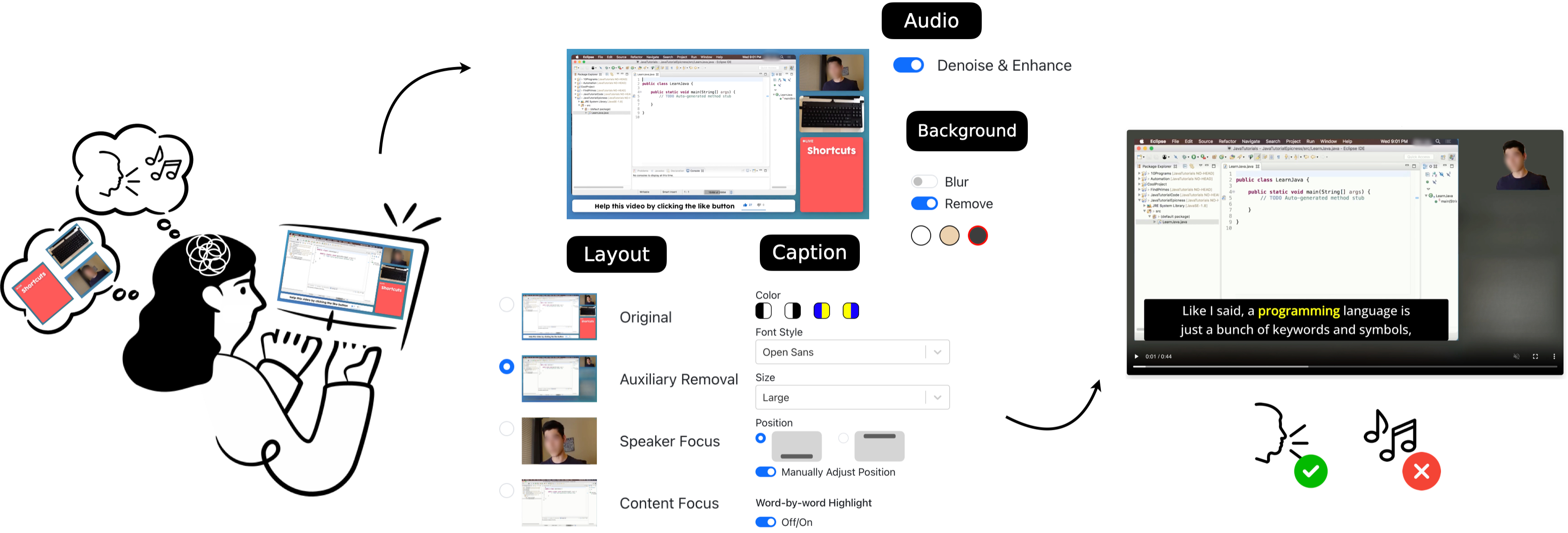}
    \caption{FocusView provides a video customization interface to help reduce distractions during video watching for viewers with ADHD. It segments a video into visual and auditory elements and allows video customizations from four aspects: (1) layouts, (2) background, (3) caption, and (4) audio.} %\yuhang{the middle column of the figure could be improved: the original video interface is too big, while the FocusView interface is barely readable. and then the result video on the right is again too small... i don't think the original video should be something to enhance---so make it much smaller}}
    \label{fig:teaser}
    \Description{Illustration of a distracted learner watching an educational video with visual and audio distractions like music and pop-ups in thought bubbles. The center shows a customizable video interface with controls for audio (denoise & enhance), background (blur/remove), layout options (original, auxiliary removal, speaker focus, content focus), and captions (color, font, size, position, highlight). A processed video on the right displays improved clarity with speaker focus and enhanced captions. Icons below indicate clear speech is kept while music is removed.}
\end{teaserfigure}


\maketitle

\section{Introduction}

% \yuhang{trim and merge first two paragraphs: people with ADHD experience challenges with consuming video contents due to (highlight the challenge at the beginning); then explain what's going on (challenge with focus, etc)... More educational videos are released online, but...} 

Video watching is an important yet challenging task for individuals with Attention Deficit Hyperactivity Disorder (ADHD), a neurodevelopmental disorder that primarily affects people's ability to pay attention and maintain concentration  \cite{doernberg2016neurodevelopmental}. While the multimodal and stimulating nature of videos can make them an effective information consumption medium for people with ADHD \cite{levenberg2023learning, vijyeta2020effect}, the dynamic visual elements and rich sound effects can also increase potential distractions \cite{shen2022understanding, boy2020audiovisual}, making videos challenging for viewers with ADHD to focus on
%challenges particularly critical for viewers with ADHD due to their increased susceptibility to distractions 
\cite{schneidt2018distraction, aboitiz2014irrelevant}. %and the difficulty restoring attention after distraction \cite{white2006training, cepeda2000task}.
Among the diverse video types, videos that serve informational purposes (e.g., educational videos) are particularly challenging as the content does not always align with viewers' interests \cite{groen2020testing}. %\yuhang{add a sentence on how informational videos are additionally challenging... the content is less interesting?}
The increasing prevalence of informational videos across different professional contexts, such as education, news, and the workspace \cite{sablic2021video, filliettaz2022video}, has further magnified this challenge for people with ADHD, who already need to put additional effort in academic and professional tasks \cite{arnold2020long, nadeau2005career, loe2007academic}. % potentially contributing to higher rates of mental health issues like anxiety and depression \cite{sciberras2009review, schatz2006adhd, d2019comorbidity}.

Prior work has started to explore the distraction challenges faced by people with ADHD when consuming visual content \cite{shaw2005impact, ben2019pay}, identifying potential distractors in reading \cite{chiorean2024adhd}, video conferencing \cite{das2021towards}, and visualization understanding \cite{tran2024discovering}. 
%Das et al. \cite{das2021towards} found that professionals with ADHD could be distracted by objects and noises in other people's backgrounds during video conferencing. Tran et al. \cite{tran2024discovering} discovered that the use of visual embellishments such as pictograms could distract people with ADHD from efficiently identifying important data from visualizations. 
Despite the growing efforts, there is a lack of research focusing on the emerging video consumption, especially the relatively tedious informational videos, exploring \textit{how the dynamic, multimodal videos affect the viewing experiences of people with ADHD}, and \textit{how to design effective technologies to enable them to better focus on video content}. To fill this gap, our research focuses on three research questions: 

\begin{itemize}
    \item RQ1: What are the key challenges that viewers with ADHD face when consuming video content?
    \item RQ2: How can we design video customization systems to enable viewers with ADHD to remove potentially distracting video elements and better focus on the key content?
    \item RQ3: What are the unique video customization preferences of viewers with ADHD that can inspire future video personalization technology?
\end{itemize}

% \begin{enumerate}
%     \item What challenges do viewers with ADHD face when watching informational videos?
%     \item What do viewers with ADHD prefer when addressing these challenges via video customization? \yuhang{trim}
% \end{enumerate}

% \yuhang{to address this question, we designed an interface to reduce distraction in the complex video presentation by leveraging the technologies, including... and allow individuals with ADHD to flexibly customize a video based on their personal preferences} 
To address these questions, we first conducted a formative study by collecting and analyzing viewer comments under ADHD-relevant videos on YouTube and TikTok to identify the video-watching challenges %---especially the video distractions (e.g., busy backgrounds, loud background music)---experienced 
faced by viewers with ADHD in the real-world context. %Based on the formative study, 
%we first conducted a motivating study to explore elements in videos that viewers with ADHD found distracting. We analyzed comments under 373 Youtube and TikTok ADHD-relevant videos, and identified categories of elements that could cause potential distraction: layouts (e.g., speakers, graphical overlays), backgrounds, captions, and audio (e.g., background music). 
%Based on the findings, 
We then designed \textit{FocusView}, a video customization interface that allows viewers with ADHD to flexibly customize the visual and audio components and reduce distractions in video viewing. By leveraging state-of-the-art computer vision and audio processing technologies (e.g., object detection \cite{shafiee2017fast}, segmentation \cite{ravi2024sam}, and inpainting models \cite{suvorov2022resolution}), FocusView breaks down video content into different channels (e.g., speaker, presentation screen, auxiliary overlays) and enables customizations in four aspects---layout, background, caption, and audio (Figure \ref{fig:teaser}). Specifically, users can choose different types of layouts with distinct focuses, blur or replace the background with different colors, customize captions by changing their parameters or dynamically highlighting the currently spoken word, and remove background music and sound effects for speech enhancement. With these options, FocusView serves as not only \textit{a video customization tool} but also \textit{a study testbed} to investigate the video viewing preferences of ADHD users. 

We evaluated FocusView with 12 participants with ADHD, who customized three short informational videos and shared their experiences and insights for video customization. Beyond short videos with consistent presentation styles, we further explored participants' customization preferences on long informational videos featuring multiple themes and video styles, investigating how they wanted to segment a long video and customize different video clips.  

% segmenting long videos to long video customization by conducting a follow-up contextual inquiry with each participant.

Our study highlighted the unique challenges faced by viewers with ADHD when watching informational videos and further %\added{For example, we found that viewers could be distracted by minor details in video presentation, such as accessories or facial features of the speaker}. 
%\yuhang{this example is too generic, do we have anything more interesting? also, use for example, xxx instead of (e.g.,xx) as this could be important content.} \added{Towards alleviating these challenges, we} \yuhang{need better transition to more important findings "then" is too plain} 
demonstrated the effectiveness of FocusView in improving their video watching experiences. Specifically, we showed that FocusView significantly improved participants' perceived video viewability by removing distracting elements and focusing on important content. Our findings further revealed participants' diverse perceptions of video distractions (e.g., background music as a distraction vs. a stimulation boost) and customization preferences (e.g., blurring the background to preserve context vs. removing the background to eliminate distractions). Additionally, we unveiled participants' different preferred strategies for long video customization (e.g., segmenting and customizing a video \textit{ad-hoc} vs. customizing all video segments before watching), highlighting their strategies in reducing customization workload in long videos (e.g., merging and applying the same edits to video segments with similar designs). Beyond the benefits of FocusView, our study also highlighted the unique challenges arising from video customization (e.g., the customization process in itself can become a new distraction) and identified the technological issues and needs in AI-based video interpretation and modification for future video customization technology. %customizing videos. 

Our contributions are threefold. First, to the best of our knowledge, this is the first study that explores and addresses video distractions for viewers with ADHD. Second, we designed and evaluated FocusView, a video customization interface that allows viewers with ADHD to customize a video to reduce distraction, thus improving video watching experiences. Third, our user study revealed the unique distraction perception and customization preferences of people with ADHD, deriving design implications to inspire future AI-powered video customization technologies for ADHD. %designs of video customization systems to make video more accessible as a source of information for viewers with ADHD.

\section{Background \& Related Work}
Our work builds on prior research that highlights the challenges faced by people with ADHD when consuming information from different modalities, the assistive technologies that support people with ADHD, and the media adaptation technologies that inspire our design. We introduce them below to contextualize our work.

\subsection{ADHD: The Challenge of Distraction in Information Acquisition} 
Attention-deficit/hyperactivity disorder (ADHD) is a neurodevelopmental disorder that affects 7.6\% of children and 6.8\% of adults \cite{Salari2023}, with inattention and/or hyperactivity/impulsivity \cite{Wilens2010-hq} being common symptoms. Exacerbated by the high rate of comorbidity (e.g., learning disabilities, dyslexia) for ADHD \cite{Sobanski2006, reale2017}, people with ADHD could face cognitive challenges in acquiring and processing information from different modalities, including text \cite{rucklidge2002neuropsychological, sexton2012co}, visuals \cite{kim2014visual, bellato2023association}, and audio \cite{blomberg2021effects, ghanizadeh2010sensory}.

Being distracted during tasks is a common challenge faced by people with ADHD when acquiring information \cite{biehl2013impact}. Distractions could come from the external environment \cite{reimer2010impact, stokes2022measuring}. For example, Ross and Randolph \cite{ross2016differences} found that students with ADHD had difficulties disengaging from distracting stimulus (e.g., television in the classroom) and returning to the original task (e.g., completing a math computation). However, distractions could also arise from poorly designed information \cite{cassuto2013using, zhang2024lingering}. For example, Tran et al. \cite{tran2024discovering} found that pictograms in visualizations could significantly slow down the response time of people with ADHD when answering visualization-related questions, as pictograms diverted their attention from the key data. These challenges highlight the need to understand and reduce distractions for people with ADHD when acquiring information.

However, little work has explored the distraction faced by ADHD viewers from the aspect of video watching. With a multimodal format, videos afford rich information with high stimulation \cite{Li2022, Wittenberg2021} and allow pausing and replaying. As a result, many individuals with ADHD prefer using videos as a source of information and learning \cite{Nikkelen2014MediaUA, borsotti2024neurodiversity}. Nonetheless, the multimodal and stimulating nature of videos also introduces potential distractions \cite{gaspar2014suppression}, such as high-tempo music and visual effects \cite{schellewald2021getting, li2024correction}. Some prior work has explored the experiences of people with ADHD when using video-related technologies \cite{levenberg2023learning, ibrahim2016synchronous}. For example, Das et al. \cite{das2021towards} conducted an interview study to examine the remote working experiences of neurodiverse professionals and found that people with ADHD could be distracted by noise and meeting backgrounds during video conferences. Jiang et al. \cite{adhdvideoaccess} explored video viewing experiences for people with ADHD via an interview study, highlighting people's frustration towards overwhelming visuals and audios during video viewing. Despite the efforts, no work has proposed technological solutions that help people with ADHD overcome these challenges during video watching. Our work seeks to fill this gap by conducting a formative study that extends prior work with more detailed video distraction categories and designing a video customization interface to help reduce these distractions.

\subsection{Assistive Technologies for ADHD}

Prior research has explored different assistive technologies for people with ADHD to manage their daily lives. Most work had focused on children with ADHD and their families \cite{stefanidi2022, silva2024}, designing situated displays \cite{Stefanidi2024}, smartwatch applications \cite{silva2023} and games \cite{sonne2016} to support health and emotion regulation for families with ADHD children. With increased attention to adult ADHD in recent years \cite{Abdelnour2022-fb}, researchers have also begun exploring assistive technologies for adults with ADHD \cite{hernandez2025decade, connell2024}. For example, Pulatova and Kim \cite{Pulatova2024} investigated the use of swarm robots to offer a customizable fidgeting experience for adults with ADHD. Savulich et al. \cite{savulich2019improvements} designed a table game for targeted cognitive training of visual attention, and showed that game-based cognitive training is an effective method for enhancing attention in adults with ADHD.

Recently, some work has explored assistive technologies to reduce distractions during tasks for people with ADHD. For example, Cuber et al. \cite{cuber2024} designed a VR studying environment with noise cancellation to help college students with ADHD focus on their school work. Lalwani et al. \cite{Lalwani2025} designed a social robot as a companion for college students with ADHD during academic tasks, showing that the presence of a social robot serves a body-doubling role---a common ADHD technique to increase focus and productivity by having someone co-present \cite{eagle2024something}. 

Despite increasing efforts to help people with ADHD overcome challenges in academic tasks, we found no work in reducing distractions in videos and improving video viewing experiences for people with ADHD, even though the video is an important source of information and knowledge input \cite{xia2022millions}. As such, our work seeks to bridge this gap by designing a system to help people with ADHD reduce distractions in videos, and use this system as a design probe to deeply understand the needs and preferences of viewers with ADHD when watching informational videos.

\subsection{Media Accessibility via Adaptation and Customization}

Recent advances in computer vision and audio technologies have enabled media content understanding and manipulation, including object recognition \cite{shafiee2017fast, ahn2015real}, object segmentation \cite{yao2020video, ravi2024sam}, object removal and inpainting \cite{chang2019vornet, Kim_2019_CVPR}, optical character recognition \cite{mori1992historical}, and audio separation and enhancement \cite{luo2018tasnet, michelsanti2021overview}. 

With these technologies, researchers have spent efforts to tailor various media to users' individual preferences, such as webpages \cite{barrett1997personalize, flesca2005mining}, multimedia documents \cite{purvis2003creating, graham1999reader}, and podcasts \cite{sailaja2024making, bbc_adaptive_podcasting}. Among different media forms, video adaptation and customization have received growing attention in recent years \cite{ram2023vidadapter}. For example, Kim et al. \cite{kim2022fitvid} designed a system that helps users adapt and customize a desktop-optimized lecture video on mobile phones by resizing and repositioning texts and graphics on lecture slides. More recently, Rajaram et al. \cite{rajaram2024blendscape} developed BlendScape, a system that allows video-conferencing participants to customize the background of their video conferences to facilitate collaboration. 

Such media adaptation techniques have also been applied to address accessibility challenges in media consumption \cite{batanero2021improving, sunkara2023enabling}. For example, Sackl et al. \cite{sackl2020} designed a desktop video player that allows low vision users to manipulate the contrast, colors, and edges of objects within videos. Natalie et al. \cite{natalie2024} created a prototype to help blind and low vision users to customize audio descriptions when watching videos. However, to the best of our knowledge, no prior work has leveraged video adaptation techniques to support the accessibility needs for ADHD viewers in video consumption.

Our research builds upon the state-of-the-art video recognition and editing techniques to develop a video customization system for people with ADHD. Our system implementation and evaluation uncover the barriers to the current AI technologies in supporting video accessibility for ADHD and derive implications for future AI-powered video customization technology.

\section{Formative Study}
%Although prior work has shown that people with ADHD could experience challenges in acquiring and comprehending visual information including visualization \cite{tran2024discovering} and texts \cite{rucklidge2002neuropsychological}, the unique challenges they encounter when watching videos---a more dynamic, richer, and multimodal form of presentation---remain unclear. 

To uncover the key challenges faced by viewers with ADHD when consuming video content and characterize the distractions they encounter, we conducted a formative study by collecting ADHD-relevant videos on TikTok and YouTube and analyzing viewers' comments under these videos to derive potential challenges presented by these videos to ADHD viewers. This approach leveraged the richness of in-the-wild videos and comments to capture viewers’ natural responses across diverse video formats. %Our findings identified key aspects of challenges experienced by viewers with ADHD during video watching, and further revealed the specific types of distractions in the multimodal videos. 

\subsection{Data Collection and Analysis}

We collected and examined the top 20 comments of more than 350 ADHD-relevant videos on TikTok and YouTube (over 7000 comments in total). To identify ADHD-relevant videos, we used keyword search (e.g., adhd) via the YouTube Data API\footnote{\url{https://developers.google.com/youtube/v3}} to collect YouTube videos, and used hashtag search (e.g., \#adhd) via a third-party scraping tool\footnote{\url{https://apify.com/clockworks/tiktok-scraper}} for TikTok. To achieve the final video dataset for each platform, we sampled videos across a diverse set of ADHD-related topics, and manually removed videos that were duplicated, non-English, or not ADHD-relevant. We then collected the top 20 comments under each video as ranked by each platform's algorithm. Among these comments, we focused on those that indicated the viewer having ADHD (e.g., if the comment mentions \textit{``my ADHD''}) to understand the video watching experiences of viewers with ADHD and their feedback on the video presentations. %\yuhang{why do we focus on comment analysis? e.g., because the comments can reflect users' experiences and feedback?} %\yuhang{how many of these comments are from people with ADHD? as this study focuses on the experience of ADHD, should only analyze the comments where the viewer self disclose their ADHD identity?}
We analyzed the comments using thematic analysis \cite{Braun2022}, %\added{with two researchers co-creating a initial codebook and three researchers iteratively coded and discussed the codes,}
focusing on viewers' challenges of watching videos and their coping strategies. %Two researchers open-coded comments under 20 randomly selected videos independently and collaboratively developed a codebook through discussions to resolve any discrepancies. We then split and coded the remaining comments. We developed themes from the codes by clustering relevant codes using axial coding and affinity diagrams. 

%\yuhang{also, based on the findings, i don't think your analysis only focused on the accessibility issues, you also talked about their coping strategies and the potential distractors in videos.}

\subsection{Findings}

We identified four video accessibility issues for ADHD: (1) prolonged video length, (2) slow pace, (3) missing captions, and (4) distracting sounds and visuals. These challenges were particularly evident for informational videos, as one YouTube comment suggested: \textit{``ADHD culture is wanting to know more about ADHD but struggling to get through a half hour informational video about it.''}  %\yuhang{do you have something to motivate for the informational videos, saying that they were particularly challenging?} 
We also found that viewers with ADHD coped with inaccessible video length, pace, and missing captions by using existing features on the video platforms, such as the \textit{Video Chapters} feature (i.e., breaking a video into sections for viewers to navigate) on YouTube \cite{youtube2025videochapters}, speed adjustment, and automatic captioning. 

However, no solutions can effectively address the distracting elements and enable viewers with ADHD to better focus on the key video contents. This gap motivated us to create an assistive tool that allows viewers with ADHD to customize videos and reduce distractions based on their personal preferences.

To inform our design, our study identified six potential distractions in videos for viewers with ADHD: 

\textbf{Speaker's Appearance:} The presence of the speaker could cause potential distraction due to their appearance (e.g., costumes or facial features) or behaviors (e.g., movement): %\yuhang{need to add a couple of sentences to explain how the speaker appearance or behavior could be distracting... quote is just evidence, cannot replace findings. the same to the rest of the points below.} 
\textit{``I was SO distracted by the green light in your glasses moving around so I had to stare at your necklace really just speaks to my ADHD.''}

\textbf{Content Overlays:} Graphical or textual overlays (e.g., animated overlays, pop-up keywords) could also distract viewers from the original content: \textit{``As a person with ADHD, the anime overlays are way too distracting!! I would rather just see your face or mouth speech patterns because it helps me focus on the content.''}

\textbf{Auxiliary Information:} Pop-up or embedded visuals not directly relevant to the video content could also distract viewers from the video: \textit{``An ad insert comes on halfway into the video and you stare blankly until you realize it's an ad. Then you realize you're not interested in the video anymore because of your ADHD.''}  

\textbf{Background Visuals:} Certain visual elements in the video background could also cause distraction: \textit{``I can't focus! The bunny in the background was too cute to be ignored for my ADHD brain!!''}

\textbf{Caption Presentations:} While many viewers with ADHD expressed a preference for captions, we also found that improper caption designs could cause distraction: \textit{``Why you put [big] captions like this... I can't watch and read this at the same time.''}

\textbf{Background Audio:} Lastly, we found that background audio (e.g., music, sound effects) could also make a video challenging for viewers with ADHD: \textit{``I have ADHD and I can't watch this because the background music is too distracting. I can't hear what you say.''}

Our video customization system for ADHD will thus cover the above six types of distractions, using computer vision and audio techniques to split the videos into multiple channels (e.g., speaker, overlays, background visuals, caption, and background audio), allowing customization and distraction reduction on these channels. % to reduce potential distraction for users with ADHD.  

%customize and reduce distractions in videos. Following the user-centered design approach, we also used FocusView as a design probe to understand \textit{how and why viewers with ADHD would like to customize a video to improve their video-watching experiences}. 

\section{FocusView: A Video Customization System for Viewers with ADHD}

Based on findings from our formative study, we designed \textit{FocusView}, an AI-powered web interface that enables viewers with ADHD to customize an informational video across different aspects (i.e., layout, background, caption, and audio). To support flexible customization of the four aspects, we utilized computer vision and audio techniques to split a video into multimodal elements, allowing viewers with ADHD to remove, keep, or focus on certain visual and auditory components based on their personal preferences. We elaborate on the feature design and implementation below.

\begin{figure*}
    \centering
    \includegraphics[width=0.85\linewidth]{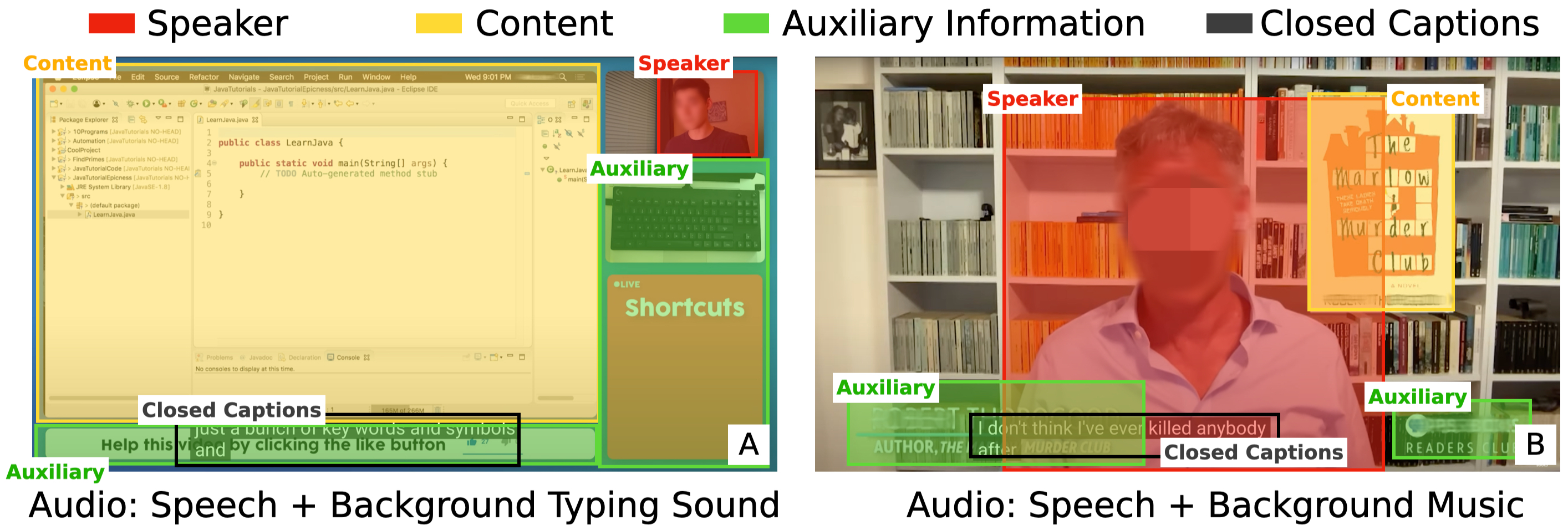}
    \caption{Illustration of video elements for two videos: speaker, content (i.e., visual elements illustrating the content being discussed, including presentation screens and graphical illustrations), auxiliary information (i.e., visual elements conveying information but not illustrating the video content), caption, and audio.}
    \label{fig:video-elements}
    \Description{Two side-by-side annotated screenshots from educational videos, labeled “A” and “B,” showing categorized video elements. Each frame includes color-coded overlays: red for “Speaker,” yellow for “Content,” green for “Auxiliary Information,” and black for “Closed Captions.” In Video A (left), a programming tutorial shows a code editor as content, a speaker in a small video frame, keyboard visuals as auxiliary info, and captions at the bottom. Audio includes speech and typing sounds. In Video B (right), a speaker discusses a book, with the book cover marked as content, background visuals labeled auxiliary, and captions at the bottom. Audio includes speech with background music. A legend at the top identifies each category color. A reviewer comment in red below the figure asks to add text labels and questions the placement of the caption label at the bottom.}
\end{figure*}

\subsection{Customization Feature Design}
\label{sub-sec:features-design}
Prior work has broken down visual elements in educational videos into \textit{speaker, screen,} and \textit{room} \cite{castillo2021production}. We adapted and expanded on this taxonomy for the more diverse, creative, and multimodal forms of informational videos online based on findings from our formative study, breaking down a video into six types of visual and auditory elements, including \textit{speaker}, \textit{content} (i.e., visual elements illustrating the core video content, such as presentation screens or graphical overlays), \textit{auxiliary} (i.e., visual elements conveying information but not illustrating the video content, such as watermarks), \textit{background}, \textit{caption}, and \textit{audio}. We illustrate the video elements in Figure \ref{fig:video-elements} using two common informational video presentations \cite{chorianopoulos2018taxonomy}.

%Following the motivating study and guided by audiovisual taxonomies \cite{bernsen1994foundations, heller2001using}
% \yuhang{what is this audiovisual taxonomies about? it's not informative enough if you only add this term and a coupld of references. How is it related to your work?} %and ADHD accessibility guidelines \cite{mcknight2010designing, guideline} 
%\yuhang{and what is this exactly?},  FocusView allows users to customize a video from four categories: \textit{\textbf{Layout, Background, Captions,}} and \textit{\textbf{Audio}}. 
% \yuhang{you had five aspects in motivating study, but changed to four aspects here... need to explain. also, i think these are just customization perspective, not categories... } %We defined each customization aspect below 
% \yuhang{you are not categorizing elements, but rather looking at the customization from different perspectives.}: 

\begin{figure*}
    \centering
    \includegraphics[width=0.85\linewidth]{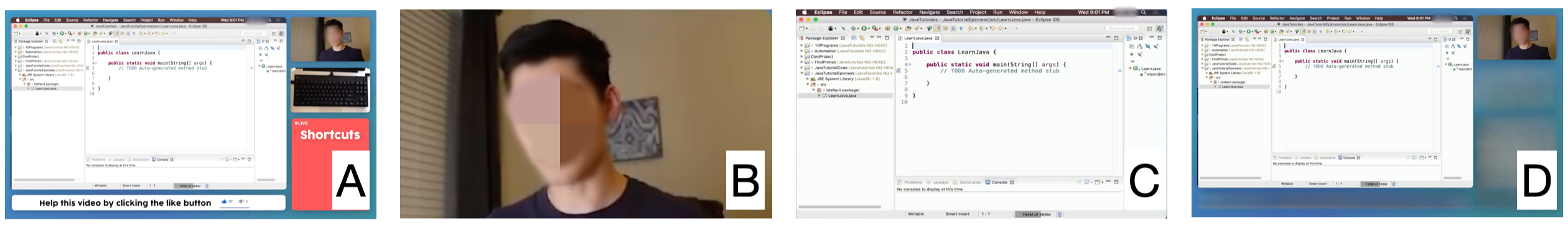}
    \caption{Layout Customization Options: (A) Original; (B) Speaker Focus; (C) Content Focus; (D) Auxiliary Removal.}
    \Description{Four side-by-side screenshots demonstrating different video layout customization options for an educational video. (A) Original: Full video frame with code editor, speaker in a corner box, and labeled shortcut panel.(B) Speaker Focus: Close-up view of the speaker only, removing all other video content. (C) Content Focus: Full-screen view of the code editor with no speaker or auxiliary panels.(D) Auxiliary Removal: Code editor remains with speaker in the corner, but the auxiliary panel is removed.}
    \label{fig:layout-options}
\end{figure*}

Based on the video elements we defined above and the formative study, FocusView allows customization in four aspects: \textit{Layout}, \textit{Background}, \textit{Caption}, and \textit{Audio}. % Targeting the background distraction identified, this customization aspect allows viewers to reduce the effect of those distractions; (3) \textbf{\textit{Caption: }} Targeting the caption design distraction identified, this customization aspect grants viewers full control of the size, style and position of the caption; and (4) \textbf{\textit{Audio: }} Targeting the background audio distraction identified, this customization aspect allows viewers to remove distracting background audio and enhance the speech.}  
We merged \textit{Speaker}, \textit{Content}, and \textit{Auxiliary Information} into the Layout aspect to offer viewers the flexibility of keeping or removing informational visual elements as they prefer. %both of them may contain important information relevant to the video content and users may have different preferences in what to remove, to keep, or to zoom into. %\yuhang{i'm just making things up here...feel free to adjust if you have better justification}. 
Moreover, to alleviate ADHD viewers' challenges with making decisions \cite{schepman2012relationship} and performing long tasks \cite{tucha2017sustained} during customization, instead of free-form customizations, we provided presets for each aspect based on the preferences of people with ADHD \cite{kolberg2020makingchoices} to ease the customization complexity. We explicate the feature design and implementation below.

\subsubsection{Layout Customization} 
%\added{As we identified both \textit{speakers} and \textit{overlays} as potential distractions in our motivating study, we also observed that viewers with ADHD might want to keep or remove those visual elements  (e.g., remove the distracting speaker vs. focus only on the speaker) for different videos. Additionally, we found that some overlays (e.g., graphical illustrations) were relevant to the video contents, while others were irrelevant (e.g., on-screen subscription messages). These differences in task relevance may influence the degree to which overlays distract viewers, with irrelevant overlays potentially being more disruptive \cite{forster2008failures, biehl2013impact}. As such, 
Based on the three potential content distractions (i.e., speaker, content overlays, auxiliary information) from our formative study, we offered three different layout simplification options (Figure \ref{fig:layout-options}) for viewers with ADHD to selectively focus on certain visual content, thus reducing distraction:
%Prior ADHD accessibility guidelines for videos and softwares \cite{mcknight2010designing, guideline} highlighted that ADHD-friendly designs should minimize distractions and irrelevant information or decorations 
% \yuhang{too generic... this is almost the motivation of the whole paper. how does it motivate the layout customization? i think you should motivate it based on the motivating study}. %Thus, we designed the layout customization feature to allow user to focus on important visual components in a video by removing irrelevant or distracting elements 
% \yuhang{this argument works for other design...background, audio, etc. try something that is specific for layout, the speaker, the content, etc.}. %We offered three different layout simplification options (Figure \ref{fig:layout-options}): 
    
\textbf{Speaker Focus:} Enlarging and centering the speaker while removing the content overlays (Figure \ref{fig:layout-options}B). We included this option as our formative study showed that some individuals with ADHD found the content overlays (e.g., pop-up graphics) distracting and preferred to focus on the speaker to leverage lip-reading to support speech comprehension, particularly due to auditory processing difficulties associated with ADHD \cite{blomberg2019speech}. 
    
\textbf{Content Focus: } Enlarging and centering the visual elements that illustrate the video content (e.g., presentation screens, pop-up graphical illustrations) while removing other elements (i.e., speaker, auxiliary overlays) to improve readability and saliency of important content \cite{guideline, mccay2012saliency} (Figure \ref{fig:layout-options}C).  %\yuhang{i'm confused... if the presentation screen and pop-ups are separated, how do you focus on them? i thought this one only focused on the presentation.} \hazel{videos we included either have the presentation or the pop-up graphics. We focused on the presentation if its available, otherwise we enlarge and center the graphics.}

\textbf{Auxiliary Removal:} Removing overlays that are not directly content-relevant (e.g., watermarks, breaking-news banners) to reduce additional distractions (Figure \ref{fig:layout-options}D).  %\yuhang{i think you should merge these implementation right after each design.} \yuhang{this description sort of conflicts with the motivating study, where you mention all kinds of overlays all together, but in your design, you handle the content-relevant pop-ups and irrelevant overlays differently; especially that the motivating study did not really mention the irrelevant overlays at all but rather talk about the more important content-related overlays.}
\subsubsection{Background Customization}
\label{sub-sub-sec:background}

\begin{figure*}
    \centering
    \includegraphics[width=0.9\linewidth]{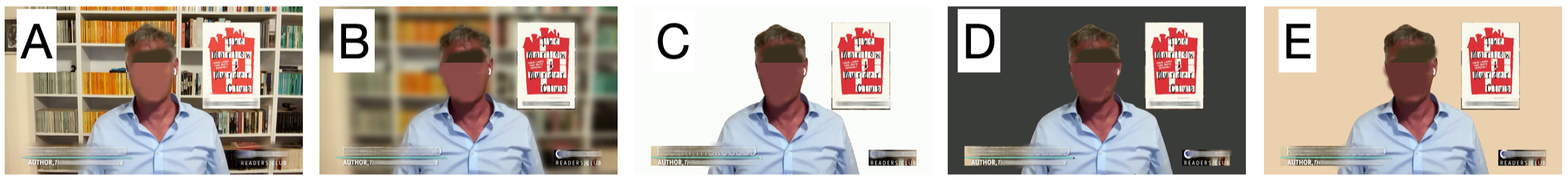}
    \caption{Background Customization: (A) Original; (B) Blur; (C) Remove (white); (D) Remove (dark); Remove (peach). }
    \Description{Five side-by-side video frames showing different background customization options for a speaker video. (A) Original: Speaker in front of a bookshelf background. (B) Blur: Original background is blurred while the speaker remains in focus. (C) Remove (white): Background replaced with solid white. (D) Remove (dark): Background replaced with solid dark gray.(E) Remove (peach): Background replaced with solid peach color.}
    \label{fig:background-options}
\end{figure*}

To reduce background distractions as highlighted in our formative study, we adapted features from common videoconferencing technologies \cite{Zoom} to offer two types of background customization: (1) \textbf{blurring} the background to reduce distraction but keep certain context (Figure \ref{fig:background-options}B); and (2) \textbf{replacing} the background with a solid color---white (\#FFFFFF), dark gray (\#3D3D3D), or peach (\#EDD1B0) (Figure \ref{fig:background-options}C-E). We used \textit{white} and \textit{dark gray} to simulate the Light/Dark theme on YouTube and \textit{peach} to offer a neurodiverse-friendly reading background \cite{rello2017}. %\yuhang{do you have a reference for ADHD instead of dyslexia?}. 

\begin{figure*}
    \centering
    \includegraphics[width=0.9\linewidth]{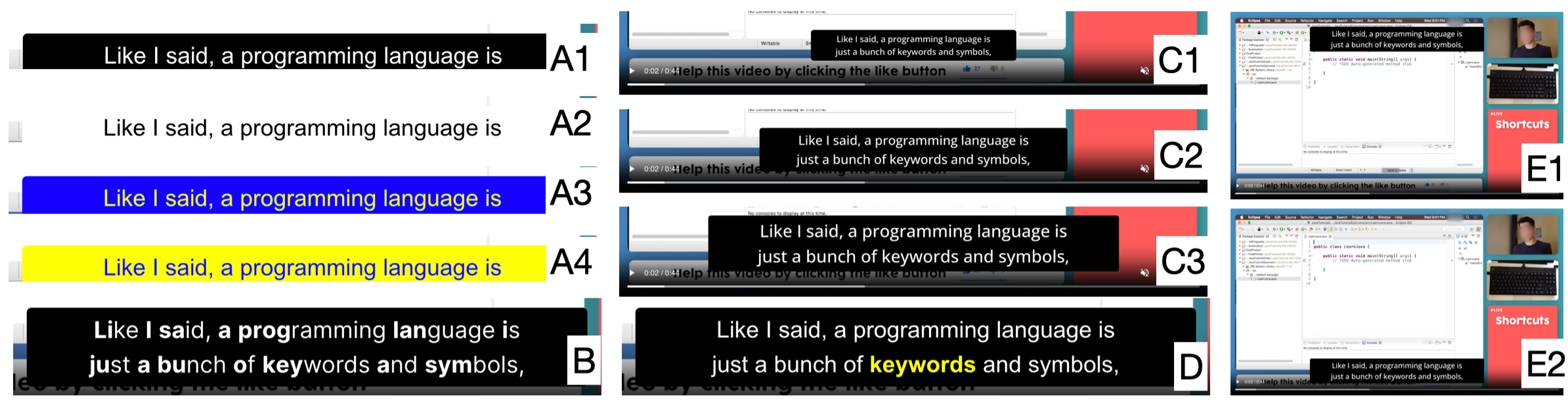}
    \caption{FocusView Caption Customization Features: (A1-A4) Color options; (B) Bionic reading font style; (C1-C3) Size options: small, medium, large; (D) Dynamic Caption Tracking; (E1-E2): Position options: top, bottom. }
    \Description{A grid of caption customization examples from a video interface labeled A1–E2. A1–A4 show different caption color backgrounds: black, white, blue, and yellow. B shows bionic reading style with bolded keyword segments. C1–C3 display font sizes: small, medium, and large. D illustrates dynamic caption tracking with real-time word highlighting. E1 and E2 compare caption positions: top and bottom of the video. Each example uses the same sentence: “Like I said, a programming language is just a bunch of keywords and symbols.”}
    \label{fig:caption-options}
\end{figure*}

\subsubsection{Caption Customization} 

Our motivating study showed that caption design could also affect ADHD viewers' video watching experiences. As such, we designed a customizable caption overlay that allows users to change the \textbf{color, font style, size,} and \textbf{position} of the caption: (1) For \textbf{color}, we offered four pairs of high-contrast font/background color combinations to alleviate potential challenges with color contrast sensitivity for users with ADHD \cite{donmez2020contrast}: \textit{white on black}, \textit{black on white}, \textit{yellow on blue}, and \textit{blue on yellow} (Figure \ref{fig:caption-options}A); (2) For \textbf{font style}, we offered Open Sans and a Bionic Reading font that is neurodiverse-friendly \cite{budomo2023impact, Born2Root} (Figure \ref{fig:caption-options}B); (3) For \textbf{font size}, we offered small (16px), medium (24px), and large (32px) options (Figure \ref{fig:caption-options}C), following the guideline that accessible font should not be smaller than 16px and users should be able to resize to 200\% \cite{section508_fonts_typography}; and (4) For \textbf{caption position}, users can toggle between the bottom and the top of the screen (Figure \ref{fig:caption-options}E), or adjust the position by manually dragging the caption on the video screen. 

To further enable users with ADHD to track and perceive the caption, we also offered a \textbf{Dynamic Caption Tracking} feature that highlights the currently spoken word in the caption (Figure \ref{fig:caption-options}D), adapted from the ADHD software accessibility guideline of highlighting alternate lines in texts to facilitate reading \cite{mcknight2010designing}. %We illustrated all caption customization options in Figure \ref{fig:caption-options}.

\subsubsection{Audio Customization} Besides visual elements, our formative study also found certain auditory elements (e.g., background music, unclear speech) could distract viewers with ADHD. Thus, we allowed users to customize the audio channel by removing background audio (e.g., music, sound effects) while enhancing speech.

\subsection{Prototype Implementation}
% short paragraph talking about models and react/js; engineering details

\begin{figure*}
    \centering
    \includegraphics[width=0.95\linewidth]{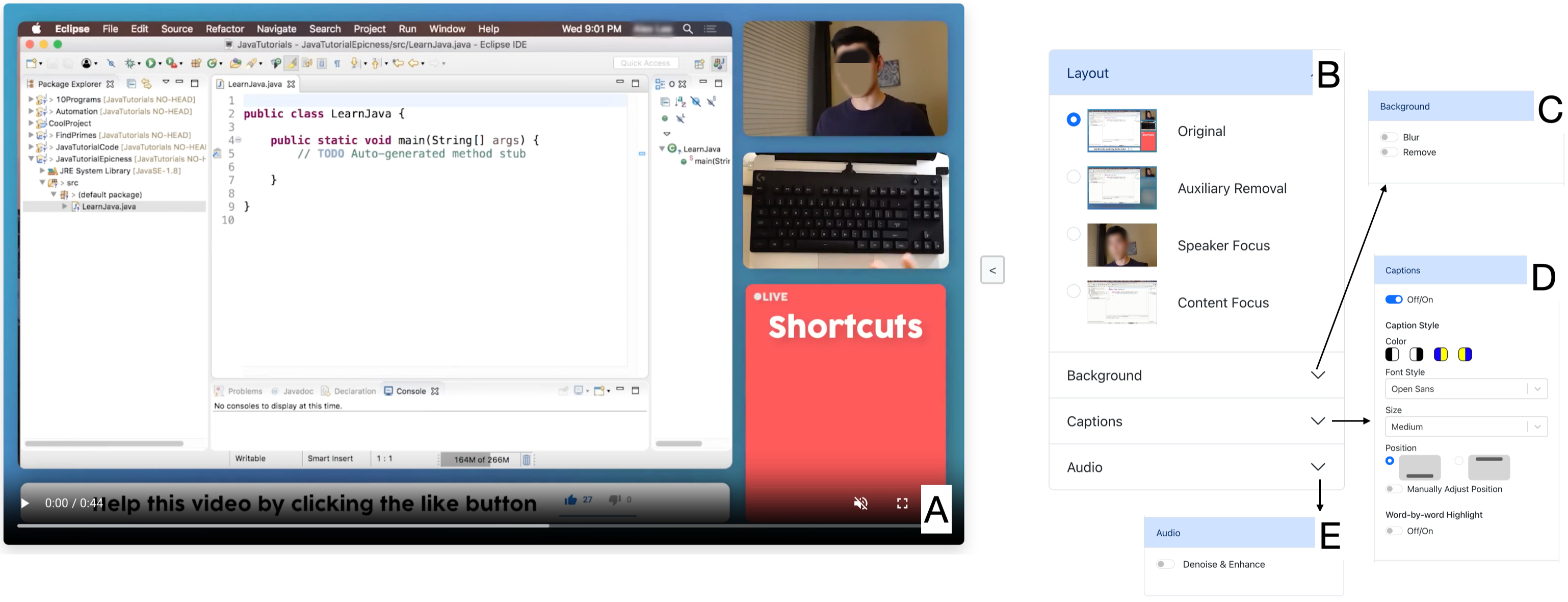}
    \caption{The interface design of FocusView, with the video player (A) on the left and the customization features (B-E) on the right. To prevent getting overwhelmed by too many options for users with ADHD \cite{kolberg2020makingchoices}, the four customization features were placed within a React accordion component to only show one feature at a time.} %\yuhang{just have one figure with all drop down menu clicked and do a screenshot}}
    \Description{A screenshot of the FocusView interface. On the left, panel (A) shows a video player displaying a coding tutorial with a visible code editor, speaker, and keyboard feed. On the right, customization options are displayed in an expandable React accordion menu with labeled sections: (B) Layout options (Original, Auxiliary Removal, Speaker Focus, Content Focus), (C) Background customization (Blur or Remove), (D) Captions settings (color, font, size, position, word highlighting), and (E) Audio enhancement (Denoise & Enhance).}
    \label{fig:interface}
\end{figure*}

We implemented FocusView by integrating all customization features into a web-based interface, as shown in Figure \ref{fig:interface}. The interface was developed with React \cite{React} and communicated with a remote server using FastAPI \cite{ramirez_fastapi} for video processing. We developed the customization features by segmenting and modifying visual and audio components through state-of-the-art computer vision and audio processing models. We elaborate on our implementation below.

\begin{figure*}
    \centering
    \includegraphics[width=0.95\linewidth]{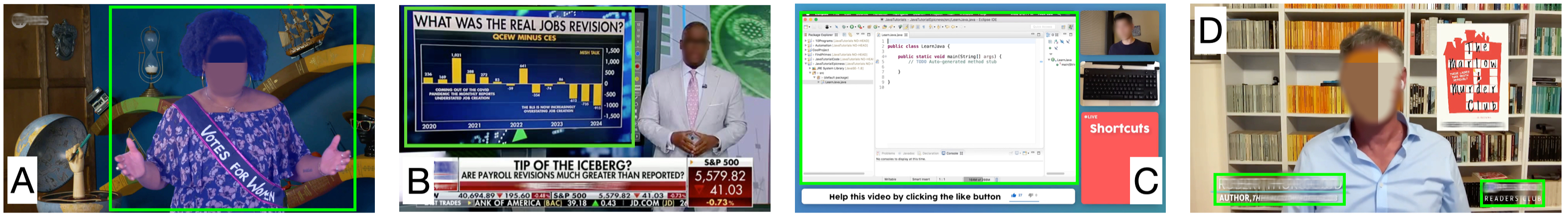}
    \caption{Examples of speaker, television and overlay recognition and segmentation. (A) Speaker; (B) Television; (C) Presentation screen; (D) Text overlay. }
    \Description{Four labeled video frames (A–D) demonstrating different types of visual elements for segmentation.(A) Speaker: A woman speaking.(B) Television: A news anchor stands beside a large TV screen displaying economic data and graphs.(C) Presentation screen: A coding tutorial showing a software interface and speaker in a side panel.(D) Text overlay: A man speaking with a title and author information displayed in green text at the bottom.}
    \label{fig:implementation-examples}
\end{figure*}

\textbf{Speaker Segmentation.} To implement the layout customization, we needed to recognize and segment various visual elements in the video, including the speaker, presentation screen, and various types of overlays. To recognize the speaker, we first used \textit{YOLO11} \cite{khanam2024yolov11} to recognize all humans in a video and then identified the speaker as the most visually salient figure based on the TranSalNet model \cite{TranSalNet}. To obtain high-quality masks for customization (e.g., background replacement), we conducted \textit{SAM2} segmentation \cite{ravi2024sam} within the speaker bounding box from the \textit{YOLO11} detection to generate more fine-grained masks (Figure \ref{fig:implementation-examples}A).%\yuhang{figure to show the implementation result, e.g., a mask within the bounding box of a speaker with labels}.

% further While YOLO could output object masks from its detection results, we wanted to obtain higher-quality masks  We thus pipelined the detection output to  to obtain a more accurate mask for the speaker.

\textbf{Presentation \& Overlays Segmentation.} By observing a large number of videos, we found that there were two common ways to present major video content: (1) through a real television, commonly in news videos (Figure \ref{fig:implementation-examples}B); and (2) via a rectangular block with text or graphics (Figure \ref{fig:implementation-examples}C). As a result, we first used \textit{YOLO11} to recognize televisions for the news context. We then recognized other types of presentations and overlays (e.g., slides, pop-up graphical illustrations) through rectangle detection. Specifically, we used Canny's Edge Detector and Probabilistic Hough Line Transform in OpenCV \cite{bradski2000opencv} to extract lines and then used rule-based methods to extract rectangular shapes among the lines by removing non-horizontal and non-vertical lines and iteratively checking whether groups of four lines form valid rectangles. We then removed rectangles with small areas (i.e., smaller than 5\% of the total video frame), and merged rectangles where one rectangle was enclosed within another. Finally, we used EasyOCR \cite{easyocr} to detect textual overlays (e.g., watermarks in text forms) (Figure \ref{fig:implementation-examples}D). 

\textbf{Layout Modification.} To enable the layout customization, we distinguished the recognized overlays into the main visual content and auxiliary based on the following rules: A detected element is a main visual content if it is (1) a television; (2) a long-term overlay that appears consistently for more than 95\% of the video duration and occupies more than 50\% of the video frame in both height and width; or (3) a large, central overlay that appears at the central area (i.e., the middle third of the frame's height) and occupies a significant portion of the frame (with either its width or height occupying at least 30\% of the frame's width or height). Any detected elements that do not fulfill the above criteria are identified as auxiliary content for potential removal in layout customization. 
We then implemented the layout customization options based on speaker detection and overlay segmentation, magnifying certain elements and removing the rest. When removing elements, we used the LaMa inpainting model \cite{suvorov2021resolution} to fill the blank. % inpainting the corresponding areas of a video frame using the .

\textbf{Background Modification.} We first segmented the background by isolating areas of a video frame excluding the speaker mask, presentation screens and overlays. We then enabled customization by blurring the background with a Gaussian Blur filter or replacing the background with a certain color using OpenCV \cite{bradski2000opencv}. 

\textbf{Caption Modification.} We generated caption files in VTT format using Whisper \cite{radford2022robustspeechrecognitionlargescale}, and then used React to overlay captions on the video player. To implement the Dynamic Caption Tracking feature, we extracted the start and end times of each speech segment from the VTT file, calculated the time per character, and estimated the duration of each word. The highlight was dynamically updated based on the elapsed time relative to each word. 

\textbf{Audio Modification.} To customize the audio, we removed all background sounds to preserve the speech component, and further enhanced the speech by restoring audio distortions and extending the audio bandwidth. Both processes were performed using the resemble-enhance model \cite{resemble-enhance}. 
\section{Evaluation}
We seek to evaluate the effectiveness of FocusView in helping viewers with ADHD better focus on informational videos, while understanding their video customization preferences. Specifically, we seek to answer the following questions:

\begin{enumerate}
    \item Whether and how does FocusView facilitate the video watching experiences of viewers with ADHD?
    \item What are the common distractions in videos, and how would viewers with ADHD like to customize them?
    \item How would viewers with ADHD like to approach customizing long videos with changing scenes? 
\end{enumerate}

\subsection{Participants}
We recruited 12 participants with ADHD (P1-12, 7 females, 3 males, 2 non-binary) whose ages ranged from 19 to 57 (\textit{Mean} = 29.7, \textit{SD} = 10.6) via email lists. All participants reported watching videos on a daily basis. Eleven participants had been clinically diagnosed with ADHD, one (P9) was in the clinical diagnostic process after being diagnosed by a psychotherapist. Table \ref{tab:users} provides participants’ detailed demographic information. A participant was eligible if they were at least 18 years old and self-reported as having ADHD. Participants were screened using a pre-study questionnaire to ensure they met these criteria. Participants were compensated \$20 per hour and reimbursed for travel expenses. This study was approved by the Institutional Review Board (IRB) at our
university. %\yuhang{need to refer to the demographic table. Also need to describe the recruitment process... maillist? contacting organizations? etc.} %We didn't exclude the experiences of participants who did not have a clinical diagnosis at the point of our study, as we recognized the challenge of obtaining ADHD diagnosis for certain population \cite{Attoe2023, Morgan2023} and would like to embrace the neurodiversity movement of bringing broader categories of mariginalized people into a solidarity \cite{graby2015neurodiversity}, following calls from prior HCI work \cite{eagle2023, das2021}. 

\begin{table*}[h]
\footnotesize 
\centering
\caption{Demographic Information of Participants.}
\begin{tabular}
{C{0.6cm}C{0.6cm}C{1.5cm}C{4.0cm}C{1.8cm}C{3.5cm}}
\toprule

\textbf{PID} & \textbf{Age} & \textbf{Gender} & \textbf{Diagnostic Status} & \textbf{ADHD Subtype} & \textbf{Commonly Used Video Watching Platforms} \\
\midrule 

P1 & 23 & Female & Clinically diagnosed at 23 & Combined & YouTube \\ \hline
P2 & 19 & Male & Clinically diagnosed at 14/15 & Hyperactive & YouTube, TikTok, Instagram \\ \hline
P3 & 34 & Female & Clinically diagnosed 6 months ago & Inattentive & YouTube, TikTok, Instagram \\ \hline
P4 & 22 & Female & Clinically diagnosed at 17 & Combined & Canvas, TV streaming services \\ \hline
P5 & 25 & Non-binary & Clinically diagnosed at 18 & Combined & YouTube, Instagram, TV streaming services \\ \hline
P6 & 32 & Male & Clinically diagnosed at 32 & Inattentive & YouTube \\ \hline
P7 & 28 & Male & Clinically diagnosed at 28 & Inattentive & YouTube, Instagram \\ \hline 
P8 & 21 & Female & Clinically diagnosed at 20 & Combined & YouTube, TikTok, Instagram, Canvas \\ \hline
P9 & 28 & Female & Diagnosed by psychotherapist at 28; currently undergoing clinical diagnosis & Unknown & YouTube, Instagram \\ \hline
P10 & 26 & Non-binary & Clinically diagnosed at 21/22 & Inattentive & YouTube, TikTok, Instagram \\ \hline
P11 & 41 & Female & Clinically diagnosed in adolescence & Inattentive & YouTube, TikTok \\ \hline
P12 & 57 & Female & Clinically diagnosed as a child & Combined & YouTube, Facebook\\

\hline
\end{tabular}
\label{tab:users}
\end{table*}

\begin{table*}[h]
\footnotesize 
\centering
\caption{Video Elements for Videos Used for Short Video Customization.}
\begin{tabular}
{C{1.5cm}C{2.5cm}C{1.5cm}C{1.5cm}C{2.3cm}C{2.5cm}}
\toprule

\textbf{VID} & \textbf{Preview} & \textbf{Speaker} & \textbf{Visual Content} & \textbf{Additional Overlays} &\textbf{Background Music/Sound} \\
\midrule 
Tutorial\footnotemark[3] & \includegraphics[width=0.9cm]{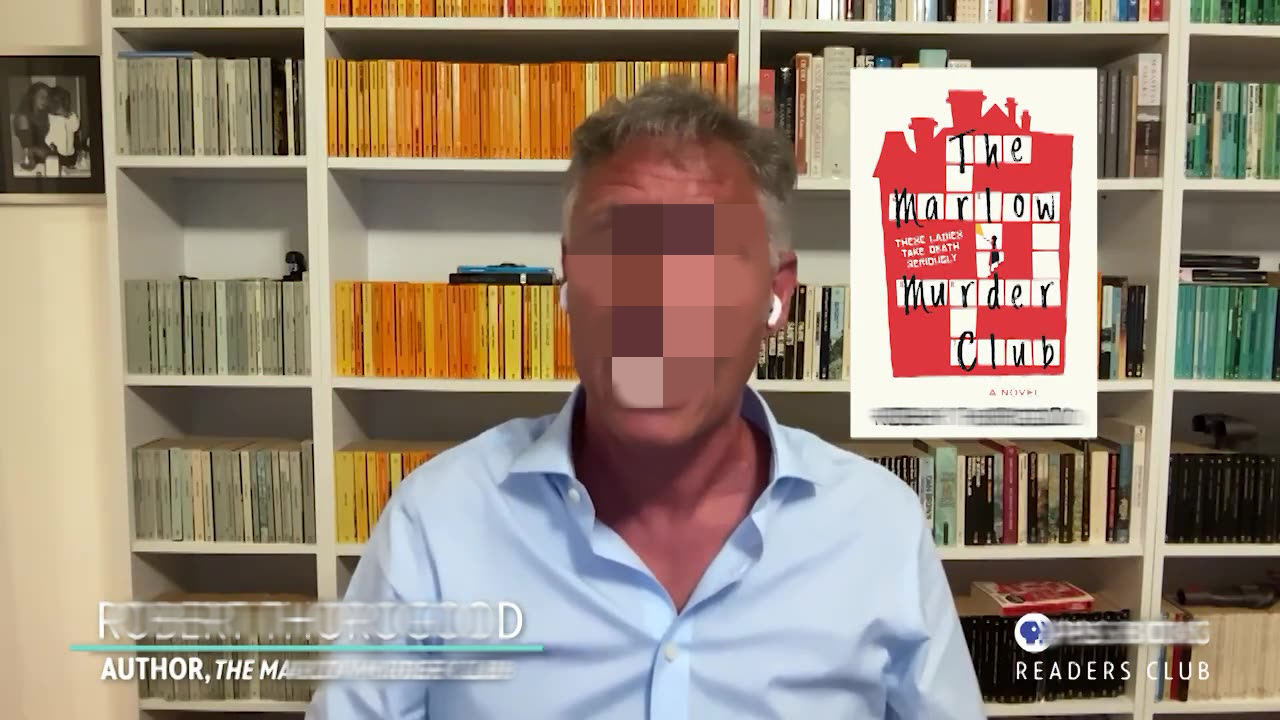} \Description{Screenshot of an talking-head video with a speaker talking and a pop-up graphic at the side.}  & \checkmark & Pop-up graphics & Watermarks &Music \\ \hline

Edu\_1\footnotemark[4] & \includegraphics[width=0.9cm]{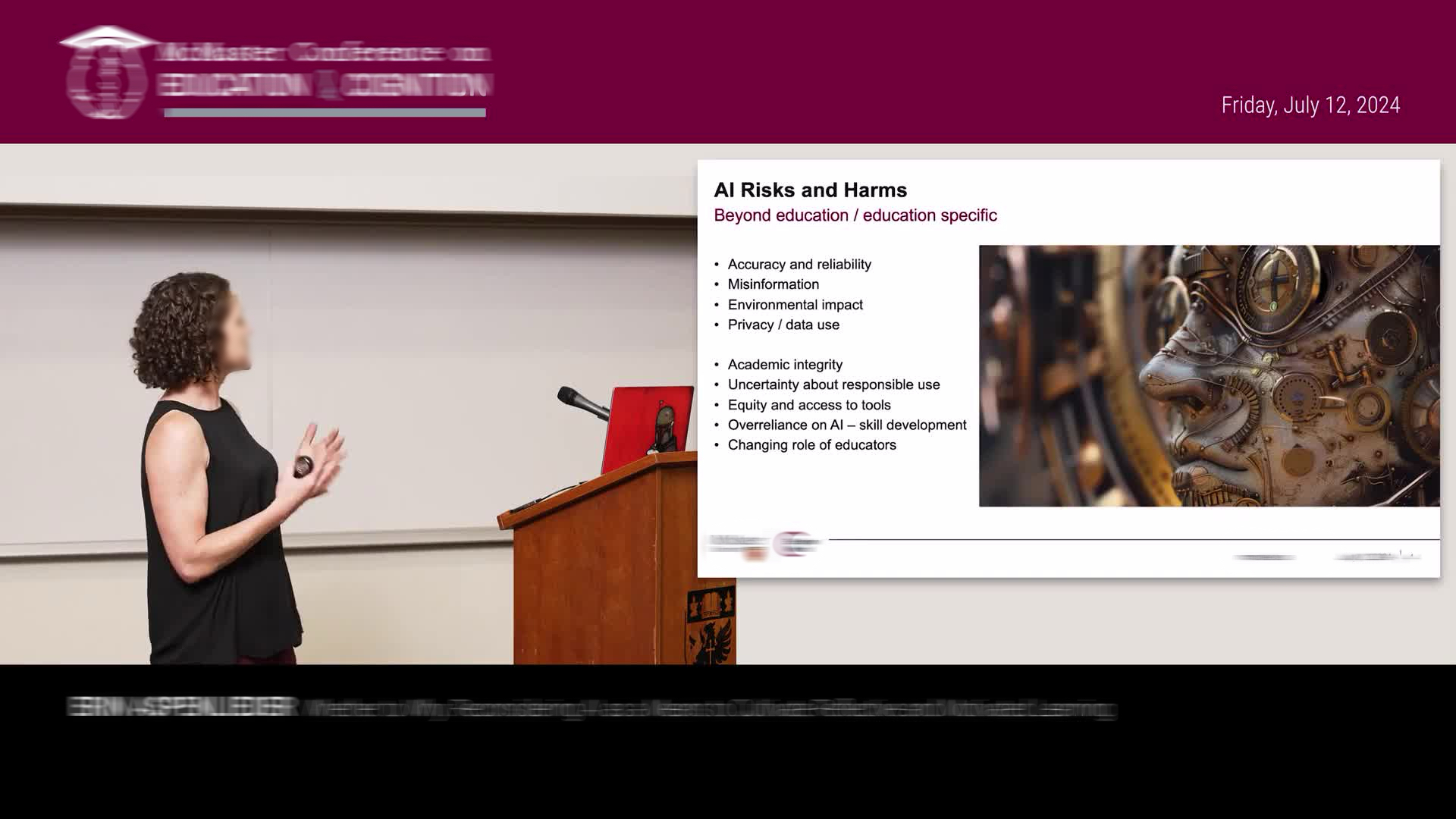} \Description{Screenshot of a speaker making a presentation with slides at the side.} & \checkmark & Screen & Banners & N/A \\ \hline
Edu\_2\footnotemark[5] & \includegraphics[width=0.9cm]{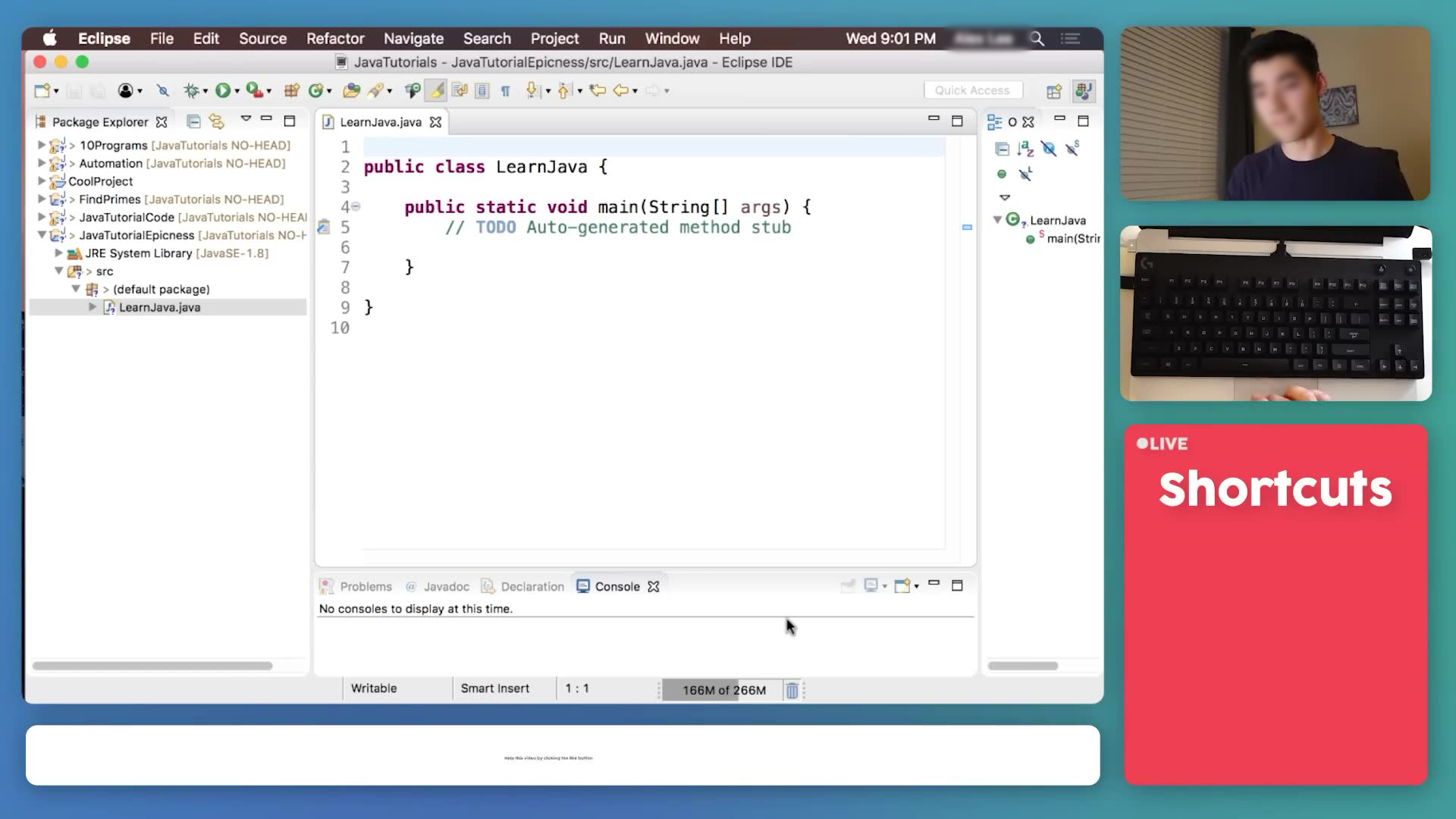} \Description{Screenshot of an educational video showing a code editor with speaker in corner.} & \checkmark & Screen & Subscreens, banners & Typing sound\\ \hline
Casual\_1\footnotemark[6]  & \includegraphics[width=0.9cm]{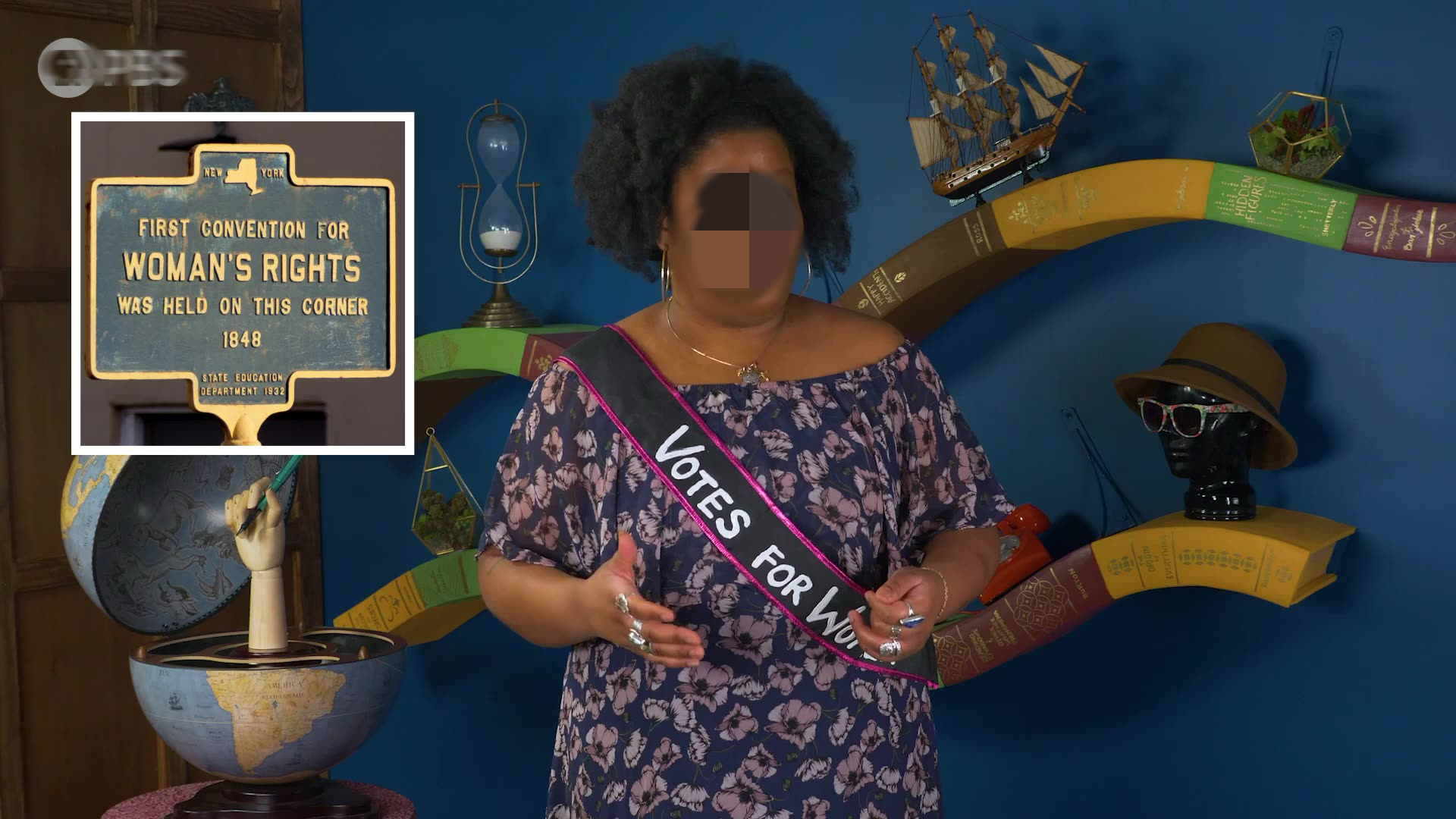}\Description{Screenshot of a talking-head video with a speaker talking and a pop-up graphic at the side.} & \checkmark & Pop-up graphics & Watermarks & Music\\ \hline
Casual\_2\footnotemark[7]  & \includegraphics[width=0.9cm]{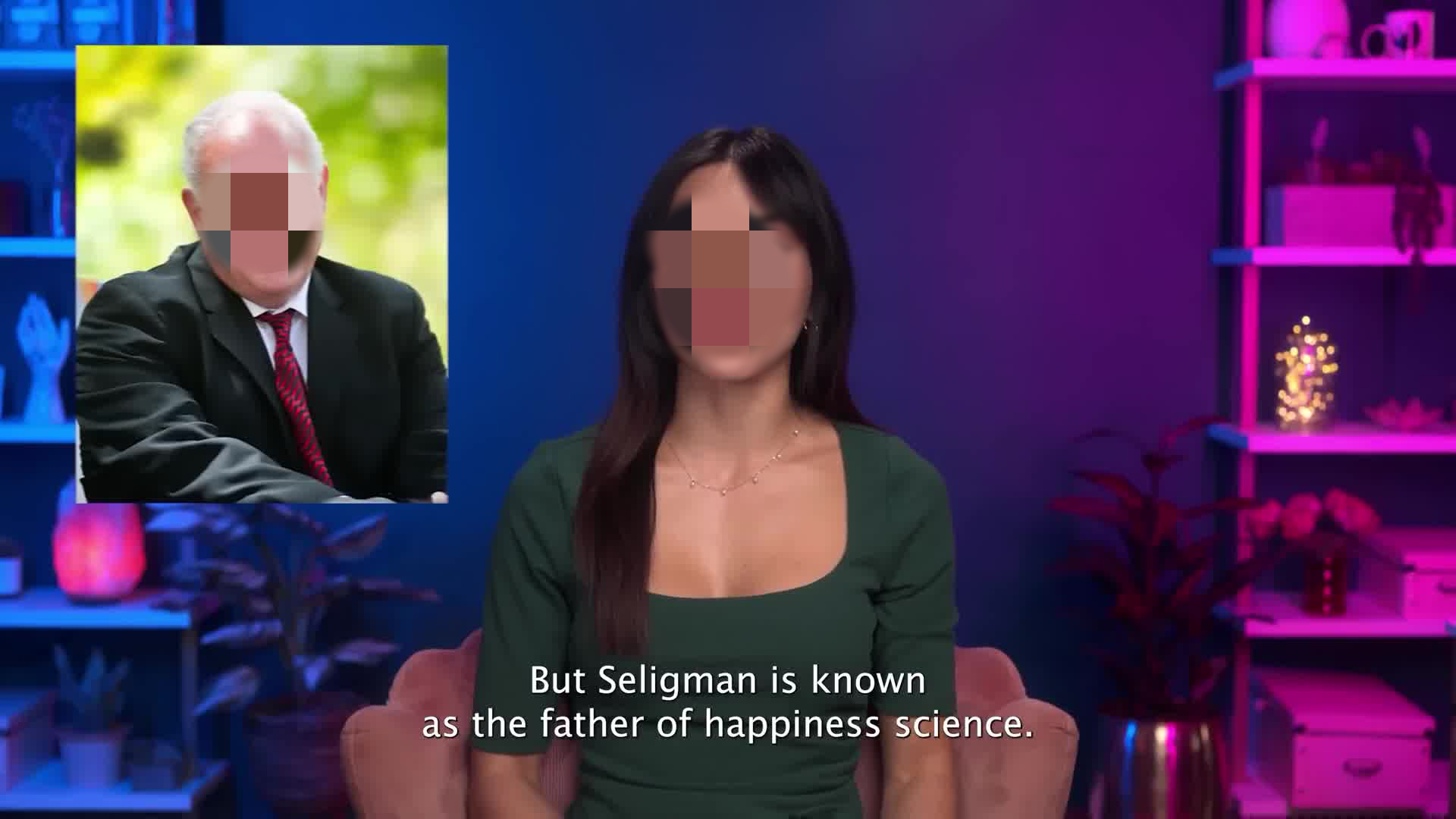} \Description{Screenshot of an talking-head video with a speaker talking and a pop-up graphic at the side.} & \checkmark & Pop-up graphics & Embedded subtitles & Music  \\ \hline
News\_1\footnotemark[8]  & \includegraphics[width=0.9cm]{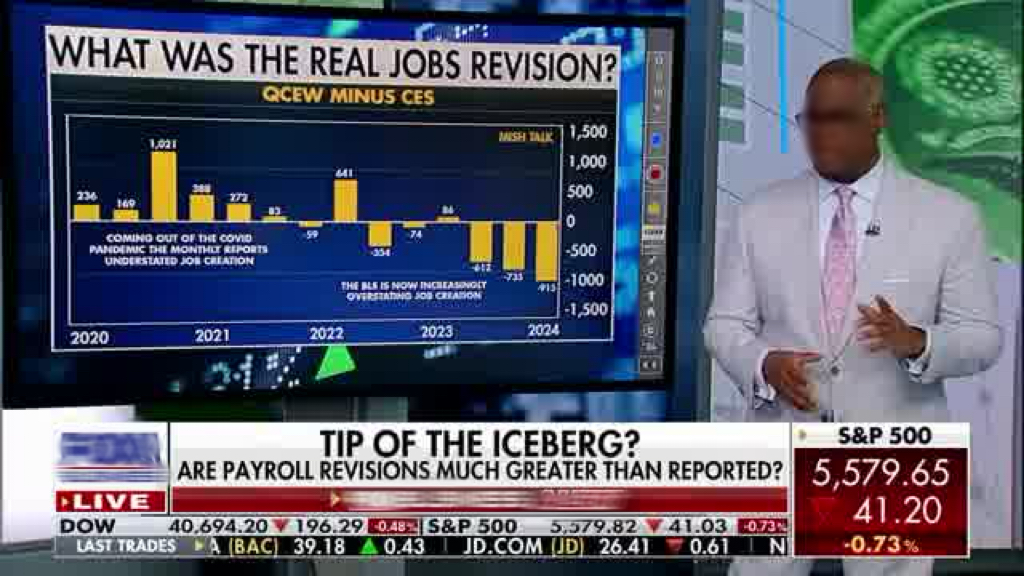} \Description{Screenshot of a news video with a news reporter explaining charts on a television.} & \checkmark & Screen & Banners & N/A \\ \hline
News\_2\footnotemark[9] & \includegraphics[width=0.9cm]{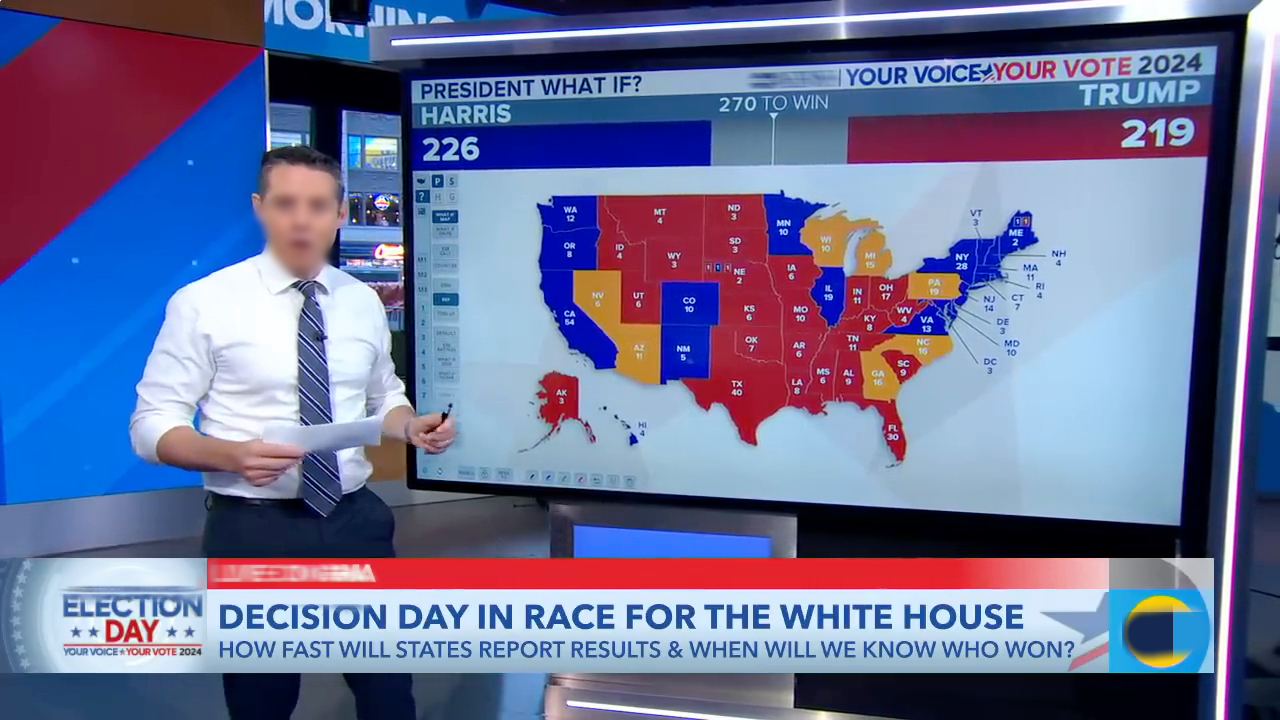} \Description{Screenshot of a news video with a news reporter explaining charts on a television.} & \checkmark & Screen & Banners & N/A \\
\hline
\end{tabular}
\label{tab:videos}
\end{table*}

\subsection{Apparatus}
\label{sub-sec:apparatus}
% To give a holistic understanding of how viewers with ADHD would like to approach video customization, we divided the study into two parts: (1) Short video customization (i.e., videos under one minute with one consistent scene) with FocusView, and (2) Interview of long video customization (i.e., three to four minutes videos with changing scenes). We explained the setup of each part below. \yuhang{the description of the two parts does not seem to belong to Apparatus.}
To investigate participants' customization processes with both short and long informational videos and ensure good coverage of diverse video types, we followed the strategies below to select videos for the user study. We elaborate on our video selection and customization interface curation details. 

\textbf{Short Video Selection and Customization Interface.} We selected six videos under three categories (two per category) to cover diverse types of informational videos \cite{bartl2018youtube, chorianopoulos2018taxonomy, kizilcec2015instructor}: (1) formal presentation-style educational videos (\textit{Edu\_1, Edu\_2}); (2) casual learning videos in a talking-head style (\textit{Casual\_1, Casual\_2}), which do not involve clear learning goals \cite{lange2019informal} and are more creative and unstructured in their presentations compared to formal educational videos \cite{nguyen2023facilitating}; and (3) news videos (\textit{News\_1, News\_2}). An additional video for tutorial purposes was selected under the casual category. %The first two categories represent common video formats within the broader domain of educational videos \cite{chorianopoulos2018taxonomy, kizilcec2015instructor}. %Along with \textit{news} videos, they constitute major types of informational content found on YouTube \yuhang{need a better reference... this youtube help link does not say much.}
We selected all videos from YouTube, the most popular video-sharing platform in the United States \cite{pew2024adults}. 
Moreover, since we aimed to understand different distractions in videos and how viewers with ADHD prefer to customize them, we selected videos with complex and diverse visual and auditory elements. We listed the multimodal elements contained in each video in Table \ref{tab:videos}. We further trimmed each video to under one minute with one consistent scene to reduce the fatigue and attention challenges associated with long tasks for users with ADHD \cite{tucha2017sustained}. 

To provide a smooth, real-time video customization experience to the users in the evaluation, we sought to avoid \textit{ad-hoc} video processing as the frame-by-frame recognition and modification of the video would take a long time. As such, we pre-processed all videos and generated all possible customization combinations (4 $\times$ 5 variations per video) on a remote server before the user study. %\yuhang{this should be moved to apparatus.}

%\yuhang{move the pre-processing of video stuff here...}

\begin{figure*}
    \centering
    \includegraphics[width=0.9\linewidth]{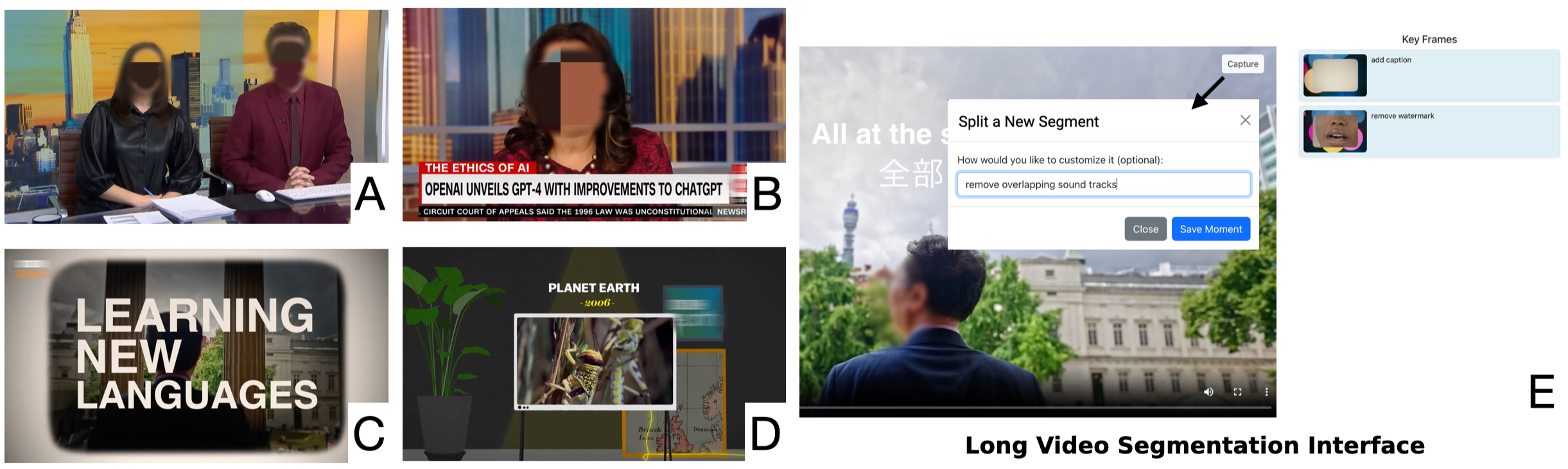}
    \caption{Videos and Interface Used in Long Video Segmentation.} %\yuhang{rename the interface moment title; and not contextual inquiry! also need to label the subimages and refer to them in the paper accordingly.}}
    \Description{Five images labeled A–E showing examples of long videos and an interface used for video segmentation. (A) A news broadcast with two anchors seated at a desk. (B) A news anchor discussing AI with a headline about GPT-4 improvements. (C) A BBC video titled “Learning New Languages” showing a silhouette of a person in front of a building. (D) A BBC Earth documentary frame showing a plant and animal-themed graphic with the title “Planet Earth - 2006 -”. (E) The FocusView segmentation interface, with a user input box titled “Split a New Segment” allowing optional customization (e.g., “remove overlapping sound tracks”), and a sidebar listing key frames labeled “add caption” and “remove watermark.” The interface allows users to select and annotate segments of longer videos for customization.}
    \label{fig:part2-videos}
\end{figure*}

\textbf{Long Video Selection and Segmentation Interface.} 
\label{sub-sub-sec:appratus-part2}
% \yuhang{i don't think this is contextual inquiry... rephrase this as an interview.}
% While we use FocusView to understand how users prefer to customize a video with diverse presentations, it is unclear whether users would like to apply the same set of customizations to a long video as its scene changes. As such, we designed the second part of the study to be an interview that began with a hypothetical task of customizing long videos with changing scenes. 
Beyond short videos, we were also interested in how viewers with ADHD would segment long videos with multiple scenes and conduct customizations accordingly. We thus designed and implemented a mock-up video segmentation interface, allowing participants to view and split a long video into segments by selecting key frames (Figure \ref{fig:part2-videos}E). The interface contained a \textit{capture} button at the top-right corner of the video player. As the participants watched a video, they could use the capture button to mark the beginning frame of a video segment that they wanted to customize, and optionally describe their preferred changes. The captured key frames would be listed beside the video player for post-watching discussions. We explain the detailed procedure in Section \ref{sub-sec:procedure}.

We selected four long videos under two categories (two videos per category): news (Figure \ref{fig:part2-videos}A\footnotemark[10] and B\footnotemark[11]) and documentary (Figure \ref{fig:part2-videos}C\footnotemark[12]and D\footnotemark[13]), as they are both informational and consist of rich variation and frequent scene switching \cite{liu2018environmental}. Each video was trimmed to three to four minutes to preserve at least 10 scene switches. %\yuhang{how many scene switching does each video consist at least? there must be some criteria for you to trim the video length. How do you define "long" compared to "short" videos? i don't think the key point is to reduce fatigue.}.

\footnotetext[3]{\url{https://www.youtube.com/watch?v=68HuYCbnhGs}}
\footnotetext[4]{\url{https://www.youtube.com/watch?v=MLvmRGzkfdU}}
\footnotetext[5]{\url{https://www.youtube.com/watch?v=RRubcjpTkks}}
\footnotetext[6]{\url{https://www.youtube.com/watch?v=kCo14j3OPTc}}
\footnotetext[7]{\url{https://www.youtube.com/watch?v=_Ta6sJmXgNs}}
\footnotetext[8]{\url{https://www.youtube.com/watch?v=YdLjEPDMT1Q}}
\footnotetext[9]{\url{https://www.youtube.com/watch?v=kvsfmJI5nJU}}
\footnotetext[10]{\url{https://www.youtube.com/watch?v=-EdgsDNwX20}}
\footnotetext[11]{\url{https://www.youtube.com/watch?v=EoJ6gXGL-zM}}

\subsection{Procedure}
\label{sub-sec:procedure}
The study consisted of a single two-hour session. We started with an initial interview, asking about participants’ demographics, ADHD background, and experiences with watching videos online. We then conducted short and long video customization sessions respectively, followed by a brainstorming session.

\footnotetext[12]{\url{https://www.youtube.com/watch?v=nzHY-muy2Mw}}
\footnotetext[13]{\url{https://www.youtube.com/watch?v=qAOKOJhzYXk}}

\textbf{Short video customization.} The goal of this session is to evaluate the effectiveness of FocusView and understand viewers' preferences in customizing videos with diverse content and presentations. %\yuhang{introduce the high-level goal first, e.g., effectiveness of the system? customization preference?}. 
We first introduced FocusView with a tutorial video (Table \ref{tab:videos}). We introduced each customization feature, asked participants to freely use them to customize the tutorial video, and collected their initial feedback on each feature. %We then invited the participants to explore the system and customize the tutorial video. 

After getting familiar with FocusView, participants watched and customized three short videos from different categories (i.e., educational, casual, news, as described in Section \ref{sub-sec:apparatus}). For each video, participants watched the first half of the video in its original form and evaluated the viewability of the video for ADHD viewers on a 7-point Likert scale. The participants then customized the video using FocusView, finished watching the customized video, and re-evaluated the viewability of the customized video. We then revisited each of their customization choices, asking about why they chose a particular option. We finally asked for additional improvements that they would like beyond the features provided by FocusView. We counterbalanced the order of the three video categories across the 12 participants. Since each video category contained two videos, we randomly assigned participants evenly across the two videos in each category, ensuring that each video was viewed by six participants. % By the end of the study, each video had been viewed six times by the 12 participants.

After watching and customizing the three short videos, participants evaluated the effectiveness, usability, and workload of using FocusView on 7-point Likert scales. We then conducted a semi-structured interview to gather their feedback on each feature, their general customization strategies, and their concerns and suggestions for using video customization systems. 

 \textbf{Long video customization.} We then conducted an interview to understand how viewers with ADHD segment and customize long, multi-scene videos. Each participant watched two long videos (one news and one documentary) and captured key frames that split a video into segments for different customizations, using the interface in Section \ref{sub-sub-sec:appratus-part2}. After watching and segmenting each video, we revisited the captured frames to discuss why they split the video at each frame, how they would like to customize each segment, and whether the customization applied beyond the segment. We counterbalanced the order of video categories and ensured that each video sample (two per category) was viewed equally across the participants. At the end of this session, we discussed participants' general preferences for segmenting and customizing long videos.

We concluded the study with a semi-structured \textbf{brainstorming session} to envision other ideas to make video watching experiences more ADHD-friendly. To facilitate ideation, we referenced previously mentioned challenges and encouraged participants to consider alternative approaches to video customization (e.g., adding or highlighting certain elements) beyond video simplification.

% \subsubsection{Video Assignment} For both parts of the study, we counterbalanced the order of the video categories (\textbf{Part 1}: educational, casual, news; \textbf{Part 2}: educational, news) viewed using Latin Square. Since each video category contained two videos, we randomly assigned half of the participants to one video and the other half to the other within each category. By the end of the study, each video had been viewed six times by the 12 participants. \yuhang{this should be merged in each phase of the study procedure.}

\subsection{Analysis}

We analyzed the study data using both quantitative and qualitative methods.

\subsubsection{Quantitative Analysis.} Our quantitative analysis focused on investigating the effectiveness of FocusView on users' video watching experience. We had one measure, \textit{Viewability}---the 7-point Likert scale scores given by participants. To evaluate the impact of FocusView on video viewability, we had one within-subject factor \textit{Condition} with two levels: \textit{WithFocusView} vs. \textit{Without}. We further investigated the impact of video types on users' viewing experiences. As such, we added another within-subject factor \textit{Video Type} with three levels: \textit{Edu}, \textit{Casual}, and \textit{News}. 

As our measure was not normally distributed based on the Shapiro-Wilk test ($p < 0.001$), we used the Aligned Rank Transform (ART) ANOVA \cite{kay2016package} to evaluate the effect of both \textit{Condition} and \textit{Video Type} on the video \textit{Viewability}. If there was any significance, we used ART contrast test
for post-hoc comparison. We used partial eta squared ($\eta_{p}^{2}$) to calculate the effect size, with 0.01, 0.06, 0.14 representing the thresholds of small, medium and large effects \cite{cohen2013statistical}.

% We checked the normality of the measure using ShapiroWilk test, and found it to \textit{not} be normally distributed. 

% \yuhang{use two-way ANOVA}
% We examined the effect of two within-subject factors on the measure. The first factor is a three-level factor named \textit{Video Type}: educational, casual, and news. The second factor is a two-level factor named \textit{Condition}: without FocusView and with FocusView. We used 

\subsubsection{Qualitative Analysis.} We audio-recorded all studies and transcribed interviews using an automatic transcription service. We analyzed the transcripts using thematic analysis \cite{braun2006using}. Two researchers open-coded three identical samples (25\%
of the data) independently and developed an initial codebook by discussing their codes. We assessed the intercoder reliability using Cohen’s Kappa ($\kappa = 0.8$), indicating substantial agreement. The two researchers then split the remaining transcripts and coded independently based on the initial codebook, periodically checking each other’s codes and discussing them to
ensure consistency. New codes were added to the codebook after the researchers reached an agreement. 

We developed themes from the codes using a combination of inductive and deductive approaches \cite{braun2006using}. Our research aims to understand different types of video distractions for viewers with ADHD, and how viewers prefer to customize a video to reduce these distractions. Thus, our high-level theme generation was guided by these objectives, following the deductive approach. Within each theme, we employed the inductive approach, generating sub-themes by clustering relevant codes using axial coding and affinity diagrams. Once initial themes and sub-themes were identified, researchers cross-referenced the original data, the codebook, and the themes to refine and finalize the code categorization.
\section{Findings}

In this section, we present both quantitative and qualitative findings on the effectiveness of FocusView, and examine how different video types influenced participants' motivation to customize videos. We further report participants’ preferences of each customization feature as well as their perspectives on customizing long informational videos. Lastly, we highlight participants' general concerns about video customization and the broader technological potentials beyond distraction removal to inform future research directions.

\subsection{Effectiveness of FocusView}

\begin{figure}
    \centering
    \includegraphics[width=0.95\linewidth]{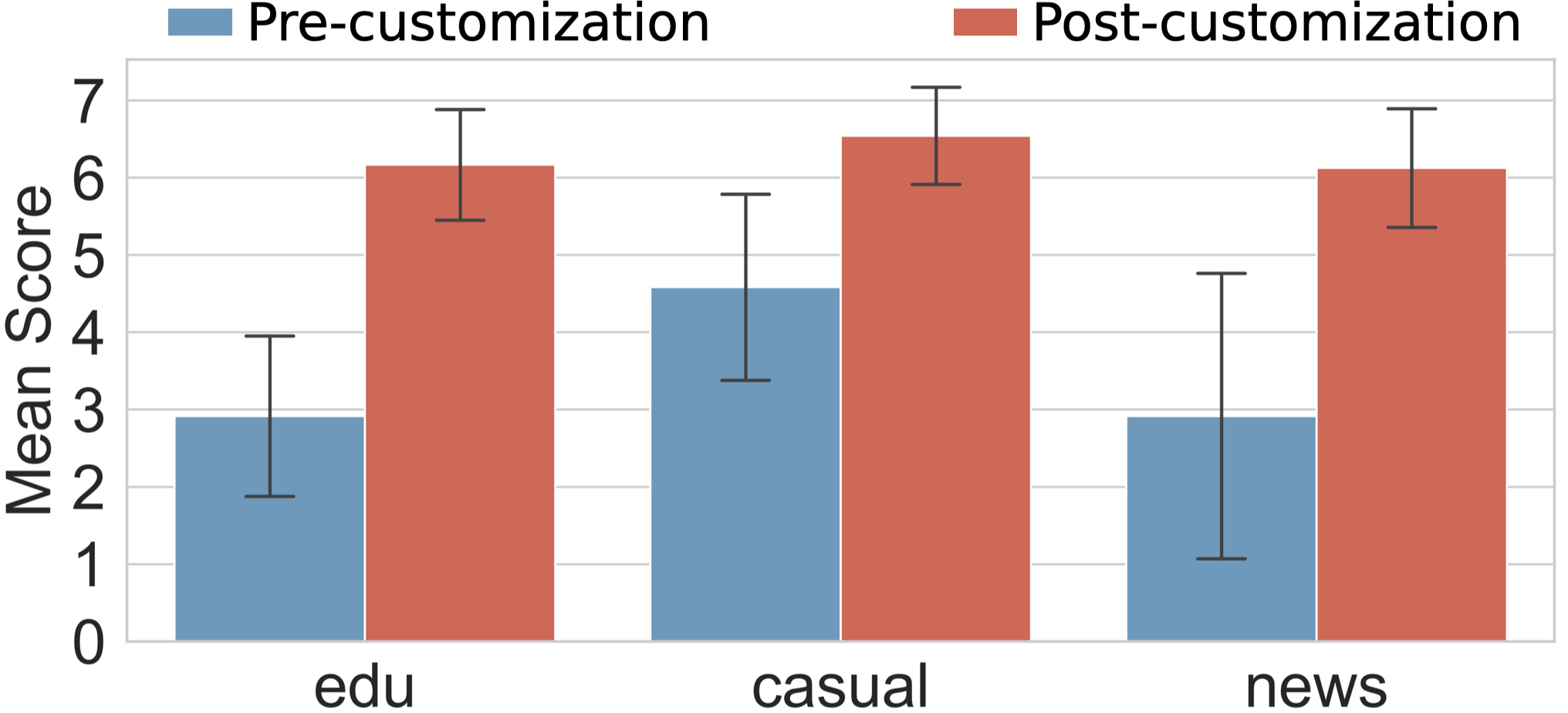}
    \caption{Likert scale comparison of video viewability for ADHD viewers before and after customization.}
    \Description{Bar chart comparing mean Likert scores for how viewable a video is for viewers with ADHD, before and after customization, across three video types: edu, casual, and news. For all three categories, post-customization scores (red bars) are higher than pre-customization scores (blue bars), indicating improved viewability. Error bars show standard deviations. Mean scores range from about 3 (pre-customization) to above 6 (post-customization) across categories, with the most improvement seen in the “edu” and “news” categories. The y-axis is labeled “Mean Score” and ranges from 0 to 7.}
    \label{fig:adhd-friendly}
\end{figure}

Participants highly rated FocusView on its effectiveness in reducing distraction and improving their video watching experiences with a mean rating of 6.17 (\textit{SD} = 0.78). They also considered FocusView easy to use (\textit{Mean} = 6.67, \textit{SD} = 0.44) with little customization effort (\textit{Mean} = 1.54, \textit{SD} = 0.66). Importantly, compared to their original video watching experience, we found that FocusView significantly improved participants' perceived viewability when watching short informational videos ($F = 165.4, p < 0.001, \eta^2_p = 0.75$). Figure \ref{fig:adhd-friendly} illustrates the comparison of the viewability scores in two conditions (with vs. without FocusView) for different types of videos, highlighting the effectiveness of FocusView across various informational video types. All participants praised the video watching experience brought by FocusView: \textit{``This is really helpful because there have been many times where I had to re-watch a video over and over again, and I wasn't getting anything out of it because it was distracting... This tool removed those distractions for me'' (P1).} P5 added: \textit{``Especially for those educational videos, I think being able to customize them would really help me focus and learn.''}%\yuhang{double check and add something about how exactly people liek it...need something qualititatvely that is very positive. quote from three participants are not enough} %Three participants (P4, P9, P12) explicitly inquired if FocusView would be an available tool for everyday use: \textit{``At first it feels like a weird concept because this isn't a thing and it feels so futuristic, but after using it... \yuhang{remove the first half, not helpful. need some other strong quotes about how exactly FocusView is helping...these pure commenting quotes are not that informative} Would this actually be a thing? Because I [would] totally use this. '' (P4).} 

\subsection{Effect of Video Characteristics on Customization Motivation}

While FocusView demonstrates its effectiveness across all informational videos, participants showed different levels of motivation to customize videos of different \textit{types}, \textit{lengths}, and \textit{styles}. Specifically, we found that participants were more willing to customize a video if it serves important purposes (e.g., for school or work), is long, or lacks strong artistic qualities. 

\subsubsection{Video Types}
As shown in Figure \ref{fig:adhd-friendly}, participants perceived \textit{casual learning} videos as the most viewable before customization. To verify this observation, we conducted another one-way ART-ANOVA test to evaluate the effect of Video Type on the perceived video Viewability \textit{before customization}, and observed a statistically significant effect ($F=9.20, p = 0.001, \eta^2_p = 0.46$). A post-hoc ART contrast test showed that \textit{casual} videos were perceived as more viewable than both \textit{educational} ($t_{22} = 3.72, p = 0.003$) and \textit{news} ($t_{22} = 3.72, p = 0.003$) videos, while no significant difference existed between the perceived viewability of educational and news videos ($t_{22} = 0.00, p = 1.00$). %We observed a statistically significant effect of video type on the perceived viewability of videos ($F = 11.7$, $p < 0.001$, $\eta^2_p = 0.30$). \added{A post-hoc analysis showed that} casual learning videos received significantly \added{higher} ratings compared to both educational \added{($estimate = 19.6, SE=4.66, t_{55} = 4.20, p < 0.001$)} \yuhang{too many parameters are reported now... please look at other paper examples for this} and news videos \added{($estimate = 19.5, SE=4.66, t_{55} = 4.19, p < 0.001$)} \yuhang{update}. However, there was no significant difference in perceived viewability between educational and news videos \added{($estimate = -0.04, SE=4.66, t_{55} = -0.01, p = 1.00$)} \yuhang{update}. 
%\yuhang{i don't think this is a correct test since what you actually want to show is that the viewability of the original videos are much higher for casual videos, and the improvement brought by FocusView was lower. However, the current test is a significant test across both conditions. So it does not address your conclusion.} %In addition, the casual learning videos have the highest average perceived ADHD-friendliness scores before customization ($Mean = 4.58 $, $SD = 1.26$)  than the other two video types (Edu: $Mean = 2.92$, $SD = 1.08 $; News: $Mean = 2.92$, $SD = 1.92$). 
Four participants (P4, P7-8, P11) highlighted that their tolerance to video distraction increased as the video became more casual: \textit{``[As this video] is not academic or news, it instantly becomes more approachable and relaxed, and I can bear with the distractions more'' (P4).} In addition, six participants (P1, P4-5, P8-9, P11) mentioned that they would be more motivated to customize academic or professionally oriented videos: 

\begin{quote}
    \textit{``If it's for class and a grade, if I need to understand something... I will definitely  try and find a setting that helps me focus... But if I'm lying in my bed watching something to relax, I'll just take [the video] as is'' (P8).}
\end{quote}

\subsubsection{Video Length} Compared to short videos, five participants (P2-3, P9-11) expressed a greater motivation to customize longer videos as distractions became more challenging with increased video length: %\yuhang{need to summarize the key points; quotes are just evidence, the paper should still stand if they are all removed. We cannot expect the readers to find out the reason by reading an individual quote.}: 
\textit{``If I watch [a] 30-second [video], I can watch it with lots of distraction if I have to. But if it's five minutes and more, the customization would really help'' (P3).} Four participants (P2-3, P9-10) mentioned 30 seconds to one minute as a length that they could manage without video customization. Two participants (P3, P9) were strongly motivated to customize videos longer than five minutes, while two (P2, P10) set the threshold to 15 to 20 minutes.

\subsubsection{Video Styles and Artistic Value.} Seven participants (P5-11) considered the style, quality, and artistic value of a video to be important factors that would affect their motivation to customize. As P7 explained, \textit{``Sometimes a video is supposed to be artistic in some way, and doing customizations affects that artistic quality... I do not want to lose that for some videos.''} Five participants (P5, P7-8, P10-11) highlighted that they would not want to customize videos with \textit{``holistic scenes'' (P5)}, such as shots of animals running in a documentary, while P6 was reluctant to customize any video that is \textit{``carefully designed.''} %P11 underscored the possibility of getting overwhelmed if they were to customize a video with artistic visuals \yuhang{the quote is not talking about artistic visuals but rather the rich and frequent changes of the visuals?}: \textit{``With all those visuals coming and so rapidly changing... I wouldn't even know how I would change it in a moment. I feel like I would be overwhelmed if I had to try to figure out how to make it better.''}

\subsection{Preferences of Video Customization}
\label{sub-sec:features}

Our study further revealed that participants with ADHD had diverse and personalized preferences for customizing videos. We explicate their preferences and strategies for each customization feature.

\subsubsection{Layout}

\begin{figure}
    \centering
    \includegraphics[width=0.95\linewidth]{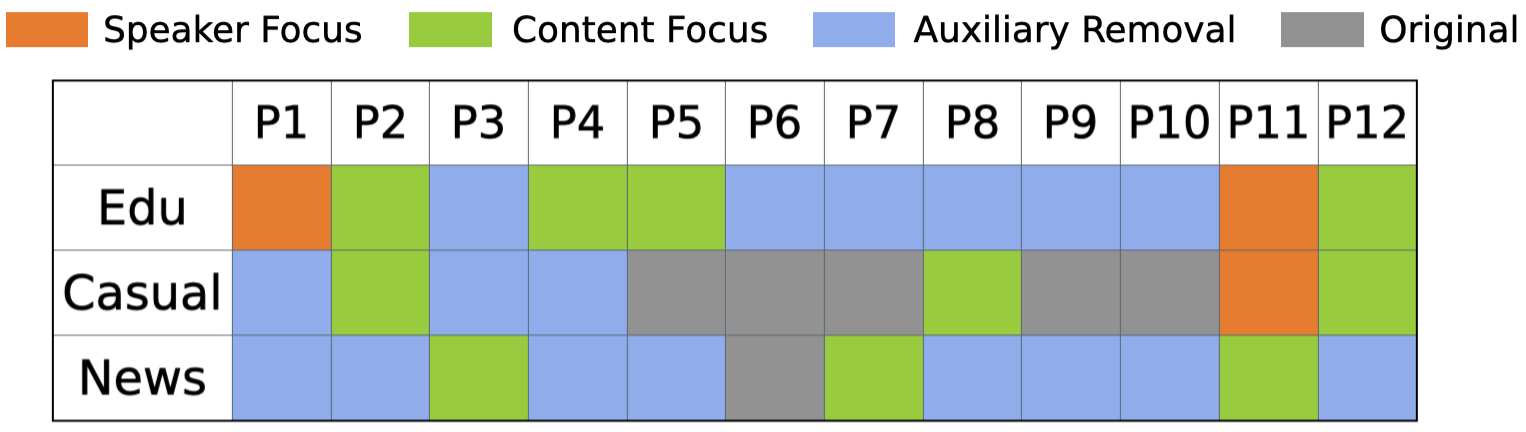}
    \caption{Participants' layout customization choices for different video types.}
    \Description{Color-coded grid showing participants’ layout customization choices across three video types: Edu, Casual, and News, for 12 participants (labeled P1–12). Each cell is color-coded based on the selected layout: Orange: Speaker Focus; Green: Content Focus; Blue: Auxiliary Removal; Gray: Original. Each row represents one video type, and each column represents one participant. Participants show varied preferences, with Auxiliary Removal and Content Focus being common choices across all video types. Speaker Focus appears most in Edu and Casual videos for participants 1 and 11.}
    \label{fig:layout-result}
\end{figure}

Participants indicated diverse preferences for video layouts, as illustrated in Figure \ref{fig:layout-result}. All participants changed their preferred layout for different videos, and no single layout was consistently selected by all participants for any video type.
%\yuhang{there's some confusion in this sentence. Do each participant had different layout preferences for different videos? or there is a difference across participants? I think you should look at both, but the chart does not seem to be effective for neither as you only had accumulated number for each video.} \yuhang{also, we cannot just simply say there is diverse preference and done... if it's diverse, you still need to describe it, e.g., how many participants changed their layout across videos, and do all participants have completely different preferences or there are some similarity or commonly used patterns. I also don't think you should do the chart based on individual videos as the two videos under each type is between participants, but different video type is within... so this chart does not really tell anything. And of course, you never just put the visualization there without explaining the findings in text.}. 
Despite diverse preferences, all participants unanimously preferred to remove irrelevant or unhelpful visual elements from a video: \textit{``Anything that's not relevant to the immediate message, not helping you understand what [the speaker] was saying, not giving additional context---all those need to go'' (P7).} Reflecting this preference, Figure \ref{fig:layout-result} showed that only 16.7\% of customization choices preserved the original layout without removing any layout element.% \yuhang{this figure does not prove this statistic at all as it's distributed across different videos, but you are reporting a total percentage.}.

%\yuhang{this section should be merged with the last paragraph... no subtitle is needed..including "feedback \& improvement." These are just generic titles, not very helpful.} %Participants considered the \textbf{relevance and helpfulness} \yuhang{i don't see a need to highlight or even summarize the finding from this perspective; relevance and helpfulness first of all is very generic, not unique findings; also, they key is that people have different preferences for different reason, i don't think the later evidence is strongly showing this relevance theme (but of course you can say that all rationales are about relevance, but that's also why it's not interesting, thus not necessary)}of a visual element to affect their layout choice. Yet, we found participants perceived these qualities in a context-dependent manner. 
Among all layout options, \textit{Auxiliary Removal} was the most frequently chosen option for both educational (50\%) and news (66.7\%) videos, and was picked by eleven participants (all except P11) during customization (Figure \ref{fig:layout-result}). Participants appreciated this option for removing the irrelevant visual information while still preserving the contexts provided by both the speaker and the main visual content: \textit{``Having [the speaker] there helps me interpret their body language and understand what's going on'' (P8). }

Interestingly, participants' preferred layout option could vary across videos. For example, P8 chose \textit{Content Focus} for the casual learning video despite choosing \textit{Auxiliary Removal} for the other two, as enlarging the pop-up graphics \textit{``really highlighted the things I need to look at.''} In contrast, P1 chose \textit{Speaker Focus} for the educational video, because the visual content (i.e., the presentation slides) \textit{``contained a lot of information,''} which could cause P1 to lose the point of focus. These diverse choices and rationales showed that participants' layout choices are highly dependent on both the video context and individual preferences.  

%\textbf{\textit{Balancing Cognitive Load and Flexibility for Layout Customization.}} 
Despite the different preferences, we found that participants (P1-3, P5, P9, P11) appreciated the limited number of layout customization options provided by FocusView. P1, P3, and P5 highlighted that having fewer layout choices reduced decision fatigue: \textit{``Another ADHD thing for me is that I am super indecisive, so having those [layout] options but not too many is good'' (P5).} Echoing this concern, P11 emphasized the importance of offering simple customization choices, as they could be distracted by the customization process if they were given too much flexibility: \textit{``If they get too complex... I would just spend all of my time messing with the [customization] settings. The settings could become a distraction.''}

However, while all participants considered the layout customization options helpful in different video contexts, seven participants (P2-4, P7, P9-10, P12) also hoped to have more flexible layout adjustments on demand. Five participants (P4, P7, P9-10, P12) desired a feature to further simplify the video by removing distracting elements (e.g., images on slides) within the presentation screen. %\yuhang{how do they want to simplify it? again, complete the sentence to make it a comprehensive finding before the quote. Should assume the writing have sufficient details without the quote.} 
For example, P4 chose \textit{Content Focus} for the educational video to more closely follow the slides. However, they still found a detailed image on the slides distracting: \textit{``I ended up mostly just focusing on this picture. It is more distracting than helpful, so if I can remove it I would''} (P4). %\yuhang{need to describe the background of this comment... have no idea what the slide is, and what is it about the picture... you do not need to always follow the same format when quoting things, e.g., we found xxx: "" (px). You can describe the story first. see my example before this quote.}. 
Moreover, three participants (P2, P3, P12) wanted to adjust the relative size of the speaker to the presentation screen in the \textit{Auxiliary Removal} mode: \textit{``It would be cool if you can decide how big the speaker is relative to their presentation, like in Zoom'' (P12).}

\subsubsection{Background}
\begin{figure}
    \centering
    \includegraphics[width=0.85\linewidth]{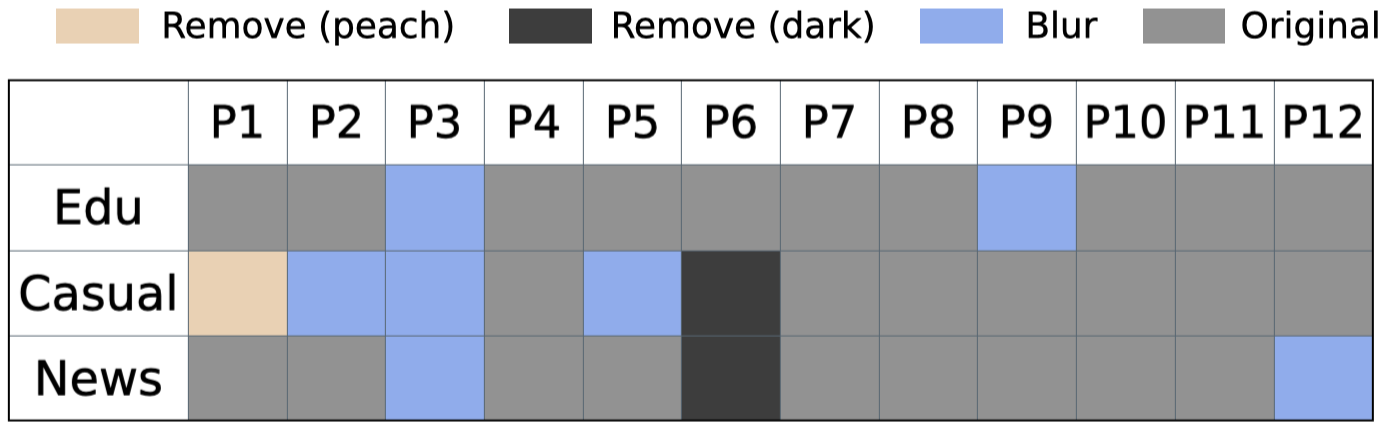}
    \caption{Participants' background customization choices for different video types. }
    \Description{Grid chart showing participants’ background customization choices across three video types—Edu, Casual, and News—for 12 participants (columns 1–12). Each cell is color-coded by background customization type: Light peach: Remove (peach); Black: Remove (dark); Light blue: Blur; Gray: Original. Most participants selected the Original background (gray) across all video types. Blur (blue) appears occasionally, especially in Casual and News. Remove (dark) is chosen by participants 6 and 7 for Casual. Remove (peach) is selected once by participant 1 for Casual.}
    \label{fig:background-result}
\end{figure}

Seven participants (P1-P3, P5-6, P9, P12) appreciated FocusView's ability to reduce background distractions: \textit{``[Background] serves as a big distraction because I will try to find literally anything else to look at other than the actual content... Having [the background customization] is definitely helpful'' (P1).} Both complex visual designs and moving background objects could serve as distractions: Eight participants (P1-3, P5-6, P9-10, P12) found the background of the casual learning video \textit{``too busy'' (P3)}, while two (P3, P12) found moving objects in the news video background (e.g., passersby of the window) distracting. Reflecting these distractions, casual learning videos had the highest rate of background customization (41.6\%), followed by news (25\%) and educational videos (16.7\%), as shown in Figure \ref{fig:background-result}.

When customizing the background, most (70\%) participant chose to blur the background rather than remove it (Figure \ref{fig:background-result}). They liked the blur option because it lessened the effect of distracting elements yet preserved the original flavor and context of the video (P3, P5, P9, P12): \textit{``[Blur] is like a depth-of-field effect... It's a natural way for [distractions] to be out of focus, but still let you know something's there'' (P5).} In contrast, two participants preferred to remove the background entirely to eliminate the distraction: \textit{``[Blur] just makes the distractions less specific... It's better to just remove them'' (P6).} P1 also appreciated having different color options. While they chose peach to remove the busy background of the casual learning video, their choice depended on their energy level: \textit{``If I was really tired, I'd choose white to brighten up my screen and make myself alerted.''} 

%Interestingly, two participants (P4, P10) complained that blurring the background could worsen the effect of distraction as it sparked their curiosity about what was being obscured: \textit{``Do you ever try to figure out what's behind the blur? I feel like I'm focused on that when it's blurred out. I always do that during zoom meetings.'' (P4).} 

%\textbf{\textit{Unintended Distractions from Background Distraction Removal.}} 
However, we also found that background customization could bring new challenges for viewers with ADHD by introducing new distractions. Seven participants (P1-3, P5, P7, P9-10) highlighted that removing the background made the boundary of the speaker appear unnatural, drawing unwanted attention: \textit{``The edges [of the speaker] become irregular... I would focus on the way her hair looks different than before'' (P5).} As a result, even though P9 and P10 considered the background of the casual learning video too complex, they didn't choose to customize it. In addition, two participants (P4, P10) also mentioned that blurring the background sparked their curiosity about what was being obscured, hence causing new distractions. 

%Additionally, three participants (P6, P9-10) also expressed a wish to keep the original background to boost their stimulation and provide more context: \textit{``It's got its own personality with the background and it fits the context. I feel like it would almost be worse if it was a plain background. I might drift away more easily if it's just a speaker floating out of nowhere.'' (P10).} 

To address this challenge, participants expressed a desire for a more natural and flexible way to simplify video backgrounds. For example, seven participants (P2, P6, P8, P9-12) wanted to remove or blur a user-defined area of distraction in the background: \textit{``[Having] a box so that we could choose what object to blur...I can just drag it to remove this boat [in the background]'' (P2).} P4 and P10 also wanted to adjust the intensity of the blur: \textit{``If it's very heavily blurred and I couldn't tell something's there at all, it would be much better'' (P10).}

%More granularity in background object removal would allow participants to preserve the original context and alleviate the challenging issue of perfecting object segmentation boundaries. %However, similar to offering flexible layout adjustment, participants (P1-2, P5, P9, P11) also mentioned their limited commitment to such a feature. \textit{``The problem is that I might not take the time to do it. I'm fairly impatient,''} said P2, after proposing the idea above, \textit{``but it's always cool to have the option.''}

\subsubsection{Caption}

Participants showed more consistent preference for captions, with nine (P1, P3-6, P8-11) turning on captions for all videos, one (P12) turning off captions for all videos, and two (P2, P7) toggling based on video content. Participants used captions both as an attention grabber and a speech clarifier: \textit{``[Captions] help me focus on the video as I read along with it... They also help clarify if I misheard something'' (P9).} Interestingly, participants toggled captions for opposite reasons: P2 turned off captions during the educational video as they found captions distracting when trying to focus on the content. In contrast, P7 only enabled captions while watching the news video with a complex chart, as captions helped them concentrate on the information being presented. This finding highlights the distinct roles that captions could play in supporting attention or leading to distraction for people with ADHD.

Participants liked the default caption design (i.e., white font on black background, Open Sans, medium size, bottom position), with two participants (P5, P8) consistently applying it to all videos. However, they also valued the flexibility to adjust caption style with FocusView. For example, P3 consistently chose the bionic reading font, finding it easier to read \textit{``for my ADHD brain''}; P9 and P10 chose the large font size for improved readability, while P11 chose the small size to \textit{``make it as unintrusive as possible.''} 

One feature that participants (P1-4, P6, P9-10) found particularly helpful was the \textit{Dynamic Caption Tracking} feature, with four participants (P1, P6, P9-10) consistently applying it to all videos. P4 explained that highlighting the currently spoken word facilitated attention recollection after distractions: \textit{``You have a deposit at any point to know where you're picking up.''} %In contrast, while P8 also found the caption highlight helpful, they wanted to highlight sentences instead of words: \textit{``I want [the caption] to be a mini-transcript with the current sentence highlighted. It gives more context. I don't need to rewind every time I feel myself missing something.''}

While some participants indicated consistent caption preferences, six participants (P1-4, P10-11) changed the caption design across different videos. For example, P1 changed the caption color to blue font on a yellow background to help them focus on the tedious news content: \textit{``It's boring, so I needed something to brighten it up and remind myself to pay attention.''} P10 manually placed the caption near the face of the speaker to allow faster attention switch: \textit{``I like [the caption] near the thing I want to focus on... I can look away at the speaker and return to [the caption] real quick'' (P10).}

\subsubsection{Audio}
\begin{figure}
    \centering
    \includegraphics[width=0.85\linewidth]{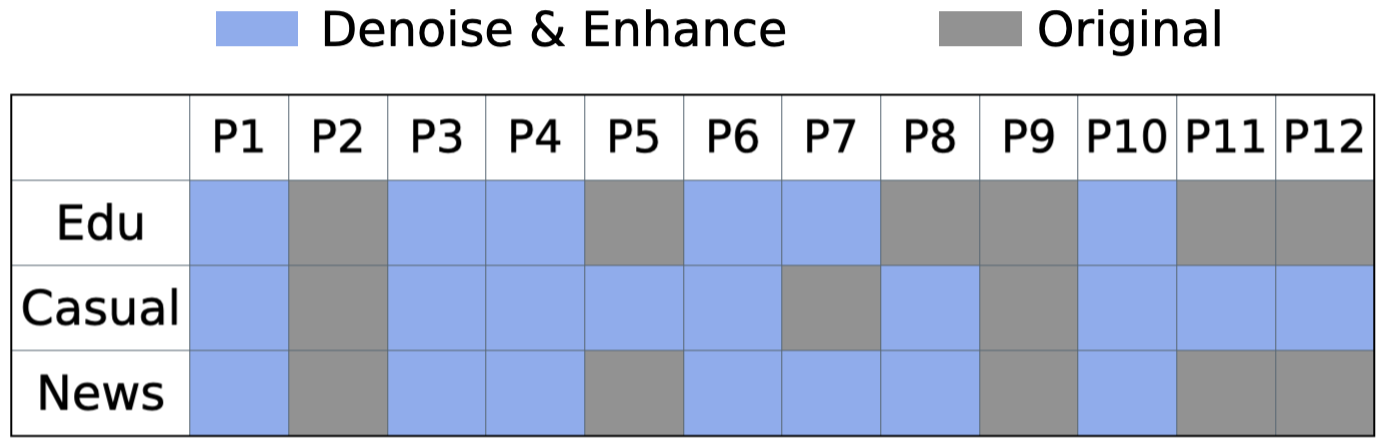}
    \caption{Participants' audio customization choices for different video types.}
    \Description{Grid chart showing participants’ audio customization choices across three video types—Edu, Casual, and News—for 12 participants (columns 1–12). Each cell is color-coded based on audio preference: Light blue: Denoise & Enhance; Gray: Original. Most participants preferred Denoise & Enhance across all video types, especially for News and Casual videos. A few participants (e.g., 2, 9) retained the Original audio in all video types.}
    \label{fig:audio-result}
\end{figure}

As shown in Figure \ref{fig:audio-result}, participants had relatively consistent preferences for audio, with five participants (P1, P3-4, P6, P10) denoising and enhancing the audio for all videos, three participants (P5, P11-12) only denoising the casual learning video that contained background music, and two (P2, P9) keeping the original audio for all videos. Three participants (P1, P3, P10) highlighted the often underestimated impact of audio distraction on viewers with ADHD, describing audio customization as the most helpful feature: 
%\begin{quote}
    \textit{``People don't realize how distracting audio can be for people with ADHD... I thought that music is going to distract me the entire time... I love this feature. I would do that for every video I watch'' (P10).}
%\end{quote}

Despite 75\% of participants choosing to remove background music in the casual learning video (Figure \ref{fig:audio-result}), we found that participants had diverse perceptions of audio as distractions. Beyond background music, four participants (P1, P3, P5, P10) also highlighted the challenge with environmental noises: \textit{``The white noises, the low humming, consistent, or high pitch sound... I can't stand them'' (P1).} In addition, certain speaking habits (e.g., lisp, vocal fry, pitch) of the speaker could also become a distraction for viewers with ADHD (P1-3). In contrast, three participants (P2, P7, P9) perceived the background audio as a stimulation boost rather than a distraction. P2 elaborated that they would add low-frequency sounds to their daily activities to enhance focus: \textit{``I can't study if it's absolute silence, so I would play four hertz frequency whenever I studied... I always prefer to have some [audio] in the background.''} 

Besides audio distraction removal, participants also appreciated FocusView's ability to enhance speech quality. This was reflected in news videos: even though without any background music or sound, seven participants (P1, P3-4, P6-8, P10) still chose to enhance the audio to improve the clarity of speech (Figure \ref{fig:audio-result}). P4 attributed their choice to their difficulty with audio processing---a common challenge faced by people with ADHD \cite{chermak1998behavioral}---noting that high quality speech could alleviate uncertainty about potential mishearing: \textit{``I know that [the speaker's] stumbling over a word than saying a word that I don't understand, which is assuring.''}

To further improve the audio customization feature, 10 participants suggested enabling more granular and intelligent control over the audio tracks being removed. Six participants (P1, P3-4, P10-12) hoped to preserve contextually relevant sounds. For example, P1 was hesitant to apply the denoise feature to a video segment when they were watching the documentary for long video customization, as the segment contained sounds from the natural environment: \textit{``The transition noise and the music are just random, but I want to keep the animal noises that are necessary for the context.''}

%Furthermore, to address the distraction from the speaker's habit of speaking, three participants (P1-3) who brought up this challenge also shared the idea of using an AI voice-over to replace the original speech: \textit{``Sometimes a video is simply frustrating to hear, and I think having this alternative hearing option would be helpful. Maybe not super monotonous, but like the typical AI-generated voice.'' (P1).} However, P3 also shared their concern towards this option from both the technical and human-centered perspectives: \textit{``The thing is that AI-generated voices are not exactly at the level. Maybe it can be, but I would probably still prefer a human voice. It's just weird to replace a human with a robot. ''}

\subsection{Customizing Long Videos}

In this section, we explain how viewers with ADHD would like to customize long videos with changing scenes.

\subsubsection{Consistency of Customization Preferences} We found that some participants maintained consistent customization preferences across scenes and videos. For example, six participants (P1, P3-4, P9-11) wanted to turn on captions at the beginning for both long videos. P1, P3 and P10 also wanted to denoise both videos to avoid any potential distraction: \textit{``I don't need the music---and I don't want any noise later that could potentially distract me from what I need'' (P3).} For news, all participants except P6 wanted to remove all news banners, logos, and pop-up advertisements throughout the video, and believed this preference applied for all news videos they watch. 

However, participants still desired flexible customization choices as the video design becomes more diverse and creative. Five participants (P3-5, P8, P11) highlighted that their preferred customization would differ for \textit{``random videos from random creators'' (P4)}. In addition, five participants (P1, P2, P4, P10, P12) wanted to apply different layouts or backgrounds during long video customization. Some of these choices can be generalized (e.g., zooming into content whenever a speaker presents a screen), while others were scene-specific: \textit{``Because of the glare from that image [in the background]... I would want to blur the background for this particular scene only'' (P10).} 

\subsubsection{Approaches to Customizing Long Videos} To achieve a balance between cognitive load and customization flexibility, eight participants (P1-2, P5, P7-11) suggested saving a preset of customization preferences and making additional adjustments for individual videos: \textit{``I really like the captions with this particular design. So maybe having that on by default, and then I can modify the rest per video as needed'' (P9).} Participants had different features that they would like to set by default. For example, P2 would like to remove auxiliary information for all lecture and news videos, while P10 would like to set the caption and denoise all videos they watched.

Participants' preferred approach for making individual video adjustments differed: nine participants (P1-3, P5, P7-9, P10, P12) wanted to make ad-hoc adjustments as they watched a video, and three (P4, P6, P11) preferred customizing individual video segments beforehand. Participants wanted to make ad-hoc adjustments to gain more contexts before removing potential distractions: \textit{``As the video plays... If I don't like something, I could turn it off and see. Rather than going in preemptively and adjusting because I would be really confused'' (P1).} On the other hand, participants who preferred pre-watching adjustments wanted a more immersive and uninterrupted video watching experience: \textit{``If I could have a preview of the different scenes and be able to modify them ahead of time, that would be nice. If I was watching [the video], I wanted to be fully watching it'' (P4).} 

\subsubsection{Splitting Long Videos}
\label{sub-sub-sec:splitting}

All participants recognized the importance of splitting a long video into segments. Even for participants who preferred making ad-hoc adjustments, five participants (P1-2, P5, P8, P10) highlighted that they would not want to automatically apply their adjustment to the rest of the video due to potential information loss: \textit{``If I zoom into the person on this [scene] and it was applied to the whole video, then I wouldn't even know the rest of the visuals existed'' (P8).} We explicate participants' preferred strategies of splitting long videos below.

\textbf{\textit{Consider Both Visual and Content Shifts. }} Three participants (P2, P5, P10) suggested video splitting based on visual scene shifts, such as \textit{``one speaker turning into two speakers'' (P5).} In addition to visual changes, two participants (P1, P9) also wanted to split a video whenever the focus of discussion changed: \textit{``If they switched to a different topic, I would like to see that [reflected] visually, and maybe change the way I customize [the video]'' (P1).}

\textbf{\textit{Exclude Irrelevant or Holistic Scenes from Customization. }} Seven participants also highlighted that video splitting should automatically exclude certain scenes (P1-2, P5, P7-8, P10-11), such as stock image inserts in news videos that were \textit{``not super relevant''} (P2), or holistic scenes such as animals running in documentaries that \textit{``felt weird to change''} (P8).

\textbf{\textit{Merge Scenes with Similar Layouts.}} Three participants (P4, P6, P11) also mentioned that they would want to avoid repeating work for scenes with similar layouts, and thus would like to merge them. For example, P11 described her preferred video segmentation after watching and segmenting the news video: \textit{``[The scenes] can be categorized: a single speaker, double speaker, and one speaker with generic stock images... These are the three main formats that exist in this video, and we can decide what we want for each format'' (P11).} %Simultaneously, P6 expressed a wish to \textit{``review''} when a customization setting was automatically applied to a differen scene: \textit{``I want to see what's there, and then I can have it simplified. I don't want to go into it without knowing.''}

\subsection{Concerns of Video Customization}

While all participants spoke positively about FocusView and the potential to customize videos, some participants also shared concerns about video customization with respect to its workload, potential waiting time, and the risks of information loss and distortion.

\textbf{\textit{Increased Workload for Long Videos.}} Eight participants (P1-2, P5-7, P9-11) expressed the concern of increased workload as videos become longer and require additional scene adjustments. P7 thought that they might only want to consistently apply caption customization for long videos, as \textit{``customizing layout for [videos] longer than five minutes would be too much work.''} Meanwhile, P11 still desired to do scene-level customization, but emphasized that the workload has to be minimal: \textit{``I don't want to feel like a film editor... I expect something that can be done within 30 seconds.''}

\textbf{\textit{Potential Video Processing Time.}} Seven participants (P4-7, P9-10, P12) raised concerns about having to wait for a video to finish processing: \textit{``I'm very, very impatient. 30 seconds is fine. Anything longer than that would be too much time'' (P12).} P10 attributed their reluctance to wait to 
their ADHD memory challenge: \textit{``With ADHD, I struggle with starting something, having to wait for it, and then remembering to come back to it. So I'll need to set an alarm to remind myself to come back, otherwise I'll just forget this thing existed.''} %This concern highlights the needs for more sophisticated parallelization techniques to process videos in customization tasks, and calls for a careful design of notifications to remind viewers upon task completion.

\textbf{\textit{Risks of Information Loss, Distortion, and New Distraction caused by AI.}} As mentioned in Section \ref{sub-sub-sec:splitting}, participants shared the concern that the automatic application of customization choices to long videos might cause unintentional or unnoticed information loss. In addition, five participants (P6-7, P9-11) also expressed the worry that AI might distort the original information presented in videos. Participants were worried that AI-generated visuals that replace video distractions could create misleading information: \textit{``There is some amount of guesswork with the AI model'' (P9).} P10 expressed the ethical concern about the \textit{``fake''} AI-generated visuals: \textit{``Even if AI could give me a really fully formed picture of what it thinks [the speaker's] legs would look like, I'd feel weird about watching that---That's not what was recorded''.} Furthermore, five participants (P4-5, P7, P9-10) also raised the concern that AI-generated visuals might introduce new visual clutters and new distractions: \textit{``Maybe [AI] will end up creating a really messy carpet pattern [to replace the distraction in video]... I would not want that'' (P7).} %\yuhang{i thought that you also had findings about the inperfect segmentation also distract the users? }

\subsection{Beyond Distraction Removal in Videos}

We designed FocusView to support video watching for viewers with ADHD mainly via distraction removal. However, we found that video accessibility for viewers with ADHD could go beyond that. We summarize participants' technological desires for improving their video watching experiences beyond distraction removal.

\textbf{\textit{Converting Video Content to Text.}} Five participants (P4, P6, P8, P9, P11) expressed a preference for accessing video content through alternative modalities, particularly text. P11 shared that for them, reading texts was often more efficient than listening to the video content: \textit{``I’m a fairly fast reader... If I’m at work watching training videos, I usually turn the volume off and read the transcript.''} While transcripts are commonly available on YouTube, P6 pointed out their usability issue: \textit{``It’s not easy to read and difficult for skimming.''} As a result, participants wanted a more reader-friendly format, including structured paragraphs with section headings. In addition to transcripts, seven participants (P1–3, P5, P7, P10, P12) proposed having video summaries to support understanding and content recall: \textit{``...a brief summary or synopsis of what the video covers at the end... to ensure that I understand what happened'' (P1).} %However, P6, P10, and P11 also expressed skepticism toward the reliability of AI-generated summaries, with P11 noting, \textit{“For transcript, it is how I’m going to get the information. But it’s always hard to say how accurate a summary can be.”}

\textbf{\textit{Adding Meaningful Illustrations.}} Three participants (P1-2, P4) suggested the idea of adding illustrations to the currently discussed content to make it \textit{``less boring''}: \textit{``Add relevant photos that are simple and associated with the content being talked about... and then leave the screen when they switched topic'' (P4).} Three other participants (P3, P5, P8) also brought up this idea, but were concerned about the potential new distractions caused by these illustrations: \textit{``... A tool might add things that are unexpected or irrelevant,  which would be more distracting than helpful'' (P5).} %Indeed, although people with ADHD could be attracted to seeking stimuli to invigorate their brains \cite{antrop2000stimulation}, having more stimuli does not always entail better engagement. P6 highlighted the distinction between attracting viewers' attention and enhancing focus on the content:

% \begin{quote}
%     \textit{``If you incorporate a lot of extra stimuli that wasn't originally there, then you're just leaning in towards this jingling thing... Like those livestreams, they are designed for attention deficits. I love them, but they're pure entertainment. It has nothing to do with engaging with the video.  You can get people to watch a video but it doesn't mean they're paying attention. ''}
% \end{quote}

\textbf{\textit{Supporting Long Video Watching.}} %As aforementioned, participants wanted some form of video summarization to help them identify interesting or important contents to watch in a video. This is similar to the video chapter feature on YouTube---five participants (P1, P3, P6, P10-11) mentioned it to be very useful in helping them watch long videos. However, instead of having a one-sentence chapter title, participants (P3, P8, P11) preferred having more details, \textit{``like a paragraph of texts'' (P8), } to give more overview on the contents being covered. 
To support information acquisition in long videos, two participants (P10–11) suggested a “bookmark” feature to capture timestamps for video moments they wanted to revisit—an idea similar to our interface for long video segmentation. As P11 explained, \textit{``I could see myself using it when I miss something, or need to go back and reinforce something I didn’t understand.''} In addition to helping viewers comprehend a video, P6 and P10 also proposed adding visual \textit{``milestones''} to track progress during video watching: \textit{``Maybe change the playback bar to a different color when a video is 25\% done.''} P10 elaborated, \textit{“You are always looking for that dopamine hit with ADHD. So if I’m watching a boring video, at least I want a sense of accomplishment.''}

\section{Discussion}
This paper explores and addresses the challenges faced by viewers with ADHD when watching informational videos. By collecting and analyzing in-the-wild ADHD-related videos and comments from YouTube and TikTok, we captured viewers’ natural responses to diverse video types and formats, and identified the multimodal distractions faced by ADHD viewers, including speakers, overlays, captions, and background visuals and audio (RQ1). Inspired by this formative study, we designed FocusView, a video customization interface that allows viewers with ADHD to customize a video from different aspects to reduce distraction and enhance focus (RQ2). We found that FocusView effectively increased the perceived viewability of informational videos for viewers with ADHD. Additionally, we examined viewers' preferences and concerns for different customization features and video types and revealed their preferred approaches to customizing long videos with changing scenes (RQ3).

In this section, we discussed the impact of our study in understanding and designing video customization systems for viewers with ADHD with concrete design implications.

\subsection{Tradeoff between Cognitive Load and Customization Flexibility}

 The potential cognitive overhead introduced by user-controlled video customization process is particularly important to consider in the ADHD context, given the cognitive challenges individuals with ADHD often face \cite{roberts2012constraints}. To address this challenge, our findings demonstrated participants' appreciation of FocusView's limited number of customization options, which helped minimize customization efforts and alleviate difficulties with decision-making. However, our work also revealed participants' desire for more flexible customization (e.g., manually defining a specific object/region to remove) even though such customization might increase cognitive load. This desire for flexibility was potentially driven by viewers' highly individualized and context-dependent preferences, which could be difficult for automated systems to satisfy. For example, while P4 wanted to preserve video background for more holistic video contexts, they sometimes wanted to remove certain background objects (e.g., a passerby in the background of an interview) that were particularly distracting to them. Our findings underscored the need to achieve a balance between reducing cognitive load and offering customization flexibility for future ADHD-oriented video customization systems.  %Future work should seek to balance cognitive load and flexible customization capability when designing video customization systems for viewers with ADHD.

Furthermore, our work also revealed additional insights for cognitive load reduction in the ADHD context. We recognized that viewers with ADHD may require an \textit{optimal} \cite{zentall1983optimal}---rather than minimal---level of stimulation to maintain focus, and video customization could serve as an added layer of stimulation when content is under-stimulating. P6 described video customization as a cognitively stimulating activity that transformed \textit{``a passive video''} into \textit{``an engaging experience,''} increasing interactivity and thus helping them focus more effectively on the video content. This sentiment aligned with the stimulation-seeking tendencies observed in some individuals with ADHD \cite{antrop2000stimulation}, underscoring the importance of helping ADHD viewers reach an optimal level of stimulation. However, P11 simultaneously underlined the risk of getting distracted by the \textit{``very fun''} customization process if it was too stimulating. As individuals with ADHD could perceive stimulations differently \cite{bijlenga2017atypical}, we encourage future work to explore customization interactions tailored to different user needs, such as allowing ADHD viewers to choose how interactive they want the customization process to be. Such flexibility would make video customization not only an accessibility tool but also an effective activity to boost focus.

\subsection{Opportunities and Concerns with AI Usage in Video Customization}

Our work revealed ADHD viewers' mixed feelings towards AI usage in video customization. On the one hand, participants recognized the value of AI to detect, segment, and remove video distractions. To further streamline the customization process, some participants also expected AI to learn their customization patterns and automatically adapt a video to suit individual preferences. Despite the potential, participants noted that current AI technologies could still present unique challenges in the ADHD context. For example, despite rapid advancement in object segmentation models \cite{ravi2024sam}, some participants still found the fuzzy boundaries of the segmented objects distracting, which led them to choose not to remove the more \textit{``natural''} video background even when they found it visually cluttered. This feedback highlighted the importance of not only focusing on distraction removal in ADHD video customization, but also considering the potential side effects, such as secondary distractions introduced by the customization itself.

Furthermore, our work uncovered the ethical concerns that ADHD viewers had towards using AI to customize videos. Some participants shared that while they liked the ability of AI to remove video distractions, they felt uncomfortable with the idea of generating new graphics to reconstruct the video scene. Echoing prior work in AI ethics \cite{xu2023combating, wach2023dark}, participants expressed concerns about potential biases (e.g., AI generating standing images of speakers who use wheelchairs), the risk of misinformation (e.g., viewers sharing customized video frames as authentic content), and the environmental cost of the computation required for video customization. As adaptive media content becomes increasingly achievable and prevalent with advances in generative AI \cite{villegas2024personalization}, we encourage future work to consider the ethical implications of these technologies, carefully evaluate the need and extent of involving generative AI in creating such tools, and incorporate protection mechanisms (e.g., AI watermarking \cite{amrit2022survey}) to combat AI misinformation.

% Our research revealed participants' mixed perceptions towards using AI for video customization. On the one hand, participants appreciated features such as using inpainting models for distraction removal and using object segmentation models to identify and blur the video background. However, we also found that AI presents unique challenges for people with ADHD. Participants mentioned that new distraction might arise when AI generates unrealistic or unnatural visuals to replace the removed distractions. Additionally, fuzzy object boundaries from object segmentation models could also cause unintended distractions, making background removal a less preferable choice in video customization for people with ADHD.

% Despite a common wish to make customized videos more natural-looking, we also found that participants were hesitant to use AI to generate realistic visuals as replacements for the original video design. Some participants were worried that AI-generated visuals might create unnecessary details that lead to new visual clutters, and some expressed concern that AI-generated content could become so realistic that it would be difficult to distinguish from the original video, potentially leading to confusion or mistrust. These concerns highlight the need to carefully design and monitor AI usage to support video customization, which we discuss in the next section. 

\subsection{Design Implications}
\label{sub-sec:design-guidelines}
Throughout the study, participants with ADHD shared their  preferences for future video customization systems. In this section, we summarize and expand on these design
implications.

\textbf{\textit{Balance Customization Workload and Flexibility. }} Throughout the study, participants expressed a wish for more targeted distraction removal and flexible adjustments, while emphasizing the challenge of increased workload that might deter them from video customization. It is thus critical to find a balance between customization flexibility and the workload involved. %As participants suggested, this could involve having a default setting to apply to videos with a particular format (e.g., news, lectures/presentations), presenting flexible adjustment options upon users' request, and having the system \textit{``learn my behavior patterns'' (P7)} to make automatic adjustments for future videos. 
Video platforms could consider having users create initial video customization profiles (e.g., removing background music for all educational videos) and presenting manual adjustment options upon users' requests. Over time, platforms could refine such profiles using ML-driven personalization techniques that learn from users’ manual adjustment history. We also encourage future research to explore more intelligent and context-aware customization solutions. For example, vision-language models (VLMs) \cite{liu2024visual} have the potential to assess whether a visual element is relevant to the video content, thus enabling automatic simplification of irrelevant elements without requiring user input.

\textbf{\textit{Limit Content and Level of Details for AI-generated Visuals.}} To mitigate the risk of AI-generated misinformation, participants highlighted that AI-generated visuals should not contain any text. Furthermore, to balance distraction reduction with information integrity, participants suggested that inpainting technologies used for distraction removal should harmonize with the rest of the video scene, but avoid generating overly detailed visuals that might attract their attention or  introduce misleading information. For example, when removing a graphical overlay, the inpainted area could use a color block that gradually blends into the video background through subtle transitions, rather than attempting to fully reconstruct the original background. While current inpainting technologies have been striving towards high fidelity \cite{li2022misf, zhang2022inertia}, we encourage future research to make careful choices when employing generative AI models for video customization purposes.

\textbf{\textit{Indicate Application of Customization and Allow Quick Information Recovery.}} To address the concern of unwanted information loss during video customization, participants suggested that future video customization systems should indicate whether and how a video has been customized. Such indicators could involve visual indications on the customized area (e.g., blur an inpainted area to distinguish it from the original video scene), or delay the automatic video customization for a few seconds to allow users to see the original information. Participants also suggested methods for quick information recovery, including hovering over an inpainted area to see the original information, or having a button to quickly switch to the original video. Future work could also consider using VLMs to summarize the customized/removed contents (e.g., ads) and list them on the side of the video in a simplified format for a quick review (e.g., text summary of an ad) without disrupting the video watching experience.
 
\textbf{\textit{Minimize Video Processing Time.}} While participants appreciated FocusView’s ability to remove video distractions through customization, they noted that their experience might differ if they had to wait for videos to process---a critical concern for people with ADHD who have shorter attention spans and higher impulsivity \cite{cuber2024}. Minimizing waiting time is thus crucial for viewers with ADHD to enjoy the video customization feature in real life. Future designs of video customization systems for viewers with ADHD should explore designs with instantaneous feedback (e.g., hover over a visual element to show a blurred overlay) for viewers to preview their customization choices, and enable parallel, progressive rendering for seamless viewing. If any waiting period is involved, the system should estimate the waiting time and proactively remind viewers (e.g., via push notifications) upon task completion. 

\subsection{Limitations and Future Work} 

Our research has several limitations. As an initial exploration aimed at supporting viewers with ADHD through video customization, our study focused primarily on understanding user needs and preferences. To eliminate potential barriers related to video processing time, we pre-processed all videos before the user study. While this allowed us to gain deeper insights into video distractions and users' customization preferences, it also limited the real-world applicability of our system. In real-world settings, processing time may present a significant challenge. Future work should address this by designing systems that account for processing delays and user experience during wait times.

Furthermore, to better understand participants' preferences for customizing videos of varying lengths and scene changes, we divided the study into two parts and selected only short videos with consistent scenes for customization via FocusView. While this design yielded valuable insights into differences in customization preferences between short and long video content, further research is needed to explore how customization systems can effectively support longer, more complex videos for viewers with ADHD. Additionally, we conducted the study in a controlled lab environment, while users' responses might differ in real-life contexts. Future work should consider deploying such systems as video platform plugins and conduct longitudinal field studies to investigate user experience and customization strategies in a real-world setting.

Finally, our overlay detection method focused on rectangular-shaped elements. In practice, many videos contain irregular-shaped illustrations or animated overlays, which require more advanced detection techniques. Future research should explore more generalizable methods for identifying and removing a broader range of distracting visual elements.

\begin{acks}
This work was partially supported by the University of Wisconsin—Madison Office of the Vice Chancellor for Research and Graduate Education with funding from the Wisconsin Alumni Research Foundation. 
\end{acks}

\bibliographystyle{ACM-Reference-Format}
\bibliography{main}


\end{document}